\documentclass[twocolumn,prx,superscriptaddress,longbibliography,nofootinbib,notitlepage]{revtex4-2}

\usepackage{graphicx}
\usepackage{multirow}
\usepackage{amsmath}
\usepackage{mathtools}
\usepackage{amssymb}
\usepackage{amsthm}
\usepackage{soul}
\usepackage{verbatim}
\usepackage{bm}
\usepackage[colorlinks=true, linkcolor=blue, citecolor=gray]{hyperref}
\usepackage[table]{xcolor}
\usepackage{tikz}
\usepackage{xspace}
\usepackage{import}
\usepackage[caption=false]{subfig}
\usepackage{bbold}

\makeatletter
\newcommand\footnoteref[1]{\protected@xdef\@thefnmark{\ref{#1}}\@footnotemark}
\makeatother

\newcommand{\ket}[1]{\ensuremath{\left\lvert{#1}\right\rangle}}
\newcommand{\bra}[1]{\ensuremath{\left\langle{#1}\right\rvert}}

\newcommand{\abs}[1]{\left|#1\right|}
\newcommand{\norm}[1]{\left\lVert#1\right\rVert}

\newcommand{\ketbra}[2]{| #1\rangle\!\langle #2|}
\newcommand{\prep}{\mathrm{PREP}}
\newcommand{\sel}{\mathrm{SEL}}

\newtheorem{theorem}{Theorem}

\newtheorem{lemma}[theorem]{Lemma}

\newcommand{\eq}[1]{(\ref{eq:#1})}
\newcommand{\thm}[1]{\hyperref[thm:#1]{Theorem~\ref*{thm:#1}}}
\newcommand{\defn}[1]{\hyperref[defn:#1]{Definition~\ref*{defn:#1}}}
\newcommand{\lem}[1]{\hyperref[lem:#1]{Lemma~\ref*{lem:#1}}}
\newcommand{\prop}[1]{\hyperref[prop:#1]{Proposition~\ref*{prop:#1}}}
\newcommand{\fig}[1]{\hyperref[fig:#1]{Figure~\ref*{fig:#1}}}
\newcommand{\tab}[1]{\hyperref[tab:#1]{Table~\ref*{tab:#1}}}
\renewcommand{\sec}[1]{\hyperref[sec:#1]{Section~\ref*{sec:#1}}}
\newcommand{\append}[1]{\hyperref[append:#1]{Appendix~\ref*{append:#1}}}
\newcommand{\cor}[1]{\hyperref[cor:#1]{Corollary~\ref*{cor:#1}}}
\newcommand{\obs}[1]{\hyperref[obs:#1]{Observation~\ref*{obs:#1}}}

\begin{document}
\title{Quantum computation of molecular structure using data from\\  challenging-to-classically-simulate nuclear magnetic resonance experiments}
\author{Thomas E.~O'Brien}
\email{teobrien@google.com}
\affiliation{Google Quantum AI, Venice, CA 90291, United States}
\author{Lev B.~Ioffe}
\email{ioffel@google.com}
\affiliation{Google Quantum AI, Venice, CA 90291, United States}
\author{Yuan Su}
\affiliation{Google Quantum AI, Venice, CA 90291, United States}
\author{David Fushman}
\affiliation{Department of Chemistry and Biochemistry, Center for Biomolecular Structure and Organization, University of Maryland, College Park, MD 20742, United States}
\author{Hartmut Neven}
\affiliation{Google Quantum AI, Venice, CA 90291, United States}
\author{Ryan Babbush}
\email{babbush@google.com}
\affiliation{Google Quantum AI, Venice, CA 90291, United States}
\author{Vadim Smelyanskiy}
\email{smelyan@google.com}
\affiliation{Google Quantum AI, Venice, CA 90291, United States}

\begin{abstract}
We propose a quantum algorithm for inferring the molecular nuclear spin Hamiltonian from time-resolved measurements of spin-spin correlators, which can be obtained via nuclear magnetic resonance (NMR). We focus on learning the anisotropic dipolar term of the Hamiltonian, which generates dynamics that are challenging-to-classically-simulate in some contexts. We demonstrate the ability to directly estimate the Jacobian and Hessian of the corresponding learning problem on a quantum computer, allowing us to learn the Hamiltonian parameters.
We develop algorithms for performing this computation on both noisy near-term and future fault-tolerant quantum computers.
We argue that the former is promising as an early beyond-classical quantum application since it only requires evolution of a local spin Hamiltonian.
We investigate the example of a protein (ubiquitin) confined in a membrane as a benchmark of our method.
We isolate small spin clusters, demonstrate the convergence of our learning algorithm on one such example, and then investigate the learnability of these clusters as we cross the ergodic to non-ergodic phase transition by suppressing the dipolar interaction. We see a clear correspondence between a drop in the multifractal dimension measured across many-body eigenstates of these clusters, and a transition in the structure of the Hessian of the learning cost-function (from degenerate to learnable). Our hope is that such quantum computations might enable the interpretation and development of new NMR techniques for analyzing molecular structure. 

\end{abstract}

\maketitle

\section{Introduction}

Quantum computing researchers are struggling to find near-term, `beyond-classical' applications of quantum computers: problems whose solution has scientific or commercial value but that cannot be solved on classical devices alone.
Fault-tolerant (FT) quantum computers able to solve valuable beyond-classical problems in chemistry~\cite{Reiher17Elucidating,Berry19Qubitization,Burg21Quantum,Lee21Even} and materials science~\cite{Babbush18Low,Kivlichan20Improved,Su2021} are predicted to be years away, leaving us in the noisy intermediate scale quantum (NISQ) era~\cite{Preskill18Quantum}.
And, while initial quantum experiments beyond classical~\cite{Google19Quantum,Wu21Strong} or on the beyond-classical boundary~\cite{Mi21Information,MI21Observation} have proven to be of great interest, it has been difficult to extend these to solve practical problems in other fields.
This is due to the large error rates on quantum computers, and to the overhead that comes from mapping non-native problems onto a quantum computer.
Here, non-native means problems beyond studying statics and dynamics of local spin systems, such as fermionic quantum simulation, linear algebra, or optimization.
Algorithms to solve these problems incur costs in circuit compilation~\cite{Whitfield11Simulation,Giovannetti08Quantum,Babbush18Encoding,Farhi14Quantum} and measurement~\cite{Wecker15Towards,Huggins19Efficient,Obrien20Nearly,Zhao20Fermionic}, creating a gap in hardware requirements before these can achieve beyond-classical results.
Before this gap is crossed, it makes sense to focus on those practical applications that do not require this overhead.

Nuclear magnetic resonance (NMR) spectroscopy has been a lauded cornerstone of analytical and organic chemistry since its discovery 80 years ago~\cite{Rabi38New,Boesch04Nobel}, being applicable to any molecule or material containing atoms with non-zero nuclear spin.
In an NMR experiment, these spinful nuclei are excited by a radio-frequency pulse and allowed to evolve for some period of time under a typically strong (multiple Tesla) magnetic field, yielding a small time-dependent response in said magnetic field.
This response, or free induction decay, contains information about its generating nuclear spin Hamiltonian, which itself contains information about the chemical structure of the molecule or material under probe.
The nuclear spin Hamiltonian is typically a strongly interacting quantum Hamiltonian due to its strong dipolar coupling.
From accurate knowledge of the dipolar couplings between different spins, one can infer the real-space molecular structure.
However, when molecules are free to quickly rotate (e.g., when in solution) this interaction averages out, leaving only chemical shifts (local fields), a weak electron-mediated Heisenberg coupling term~\cite{Vandersypen04NMR}, and residual incoherent (classical) processes.
Such Hamiltonians can be easily classically analysed (sometimes even by intuition), one of the reasons why NMR has shown such success to this day.
However, NMR spectra are considerably more complex for systems that are not free to tumble in all directions (as they are in a solution field~\cite{Mcdermott12solid}): e.g., solid-state materials~\cite{Sakellariou00Homonuclear,Madhu09High}, molecules in gels~\cite{Nonappa16Solid}, stuck on surfaces~\cite{Deng08Solid} or in membranes~\cite{Agarwal14Denovo,Sinnaeve20Selective}.
Spectrum prediction can also pose a challenge for experiments operating in low magnetic field such as `zero-field NMR' experiments~\cite{Ledbetter09Optical,Theis11Parahydrogen,Blanchard13High,Barskiy19Zero}, which promise more affordable and practical NMR technologies not requiring huge magnetic fields.
These systems present a region of parameter space where data cannot be analysed by classical computers, yielding a potential area for beyond-classical quantum computation.
This ability to analyze classically intractable data sets could enable new types of NMR techniques to characterize previously difficult-to-characterize systems.

Quantum dynamics are generated by a system's Hamiltonian; from a sufficient set of experimental data, it should be possible to learn the Hamiltonian that generated it.
Hamiltonian learning is the inverse problem to predicting experimental outcomes given a system's Hamiltonian~\cite{DaSilva11Practical}, and is well-established in the field of quantum information as an approach to device characterization.
Sufficiently-small devices may be characterized via classical post-processing of experimental data via Bayesian~\cite{Granade12Robust, Sergeevich12Optimizing} or machine-learning methods~\cite{Valenti21Scalable,Bienias21Meta,Gentile21Learning}.
Exact classical methods are intractable in large systems whenever the forward problem becomes beyond-classical, but this can be avoided when possible by careful experiment design. For example, given sufficient control one can dynamically decouple a small subsystem from its environment, even in the presence of large background magnetic fields, and then learn the global structure piece-by-piece~\cite{Somma08Parameter,Ajoy13Quantum,Wang15Hamiltonian}.
It is also possible to learn Hamiltonians from expectation values of thermal or long-time average states, which are more easy to classically approximate~\cite{Bairey19Learning,Evans19Scalable}.
These methods also allow one to learn a Hamiltonian from highly accurate measurements at ultra-short periods of time, where $e^{iHt}\simeq 1+iHt$ is a nearly-exact approximation.
However, when none of the above methods are viable, the Hamiltonian learning problem becomes classically challenging, giving a potential beyond-classical quantum computing application.
This still requires that the experiment yield sufficient data that the Hamiltonian can be inferred.

If a quantum system is chaotic, or ergodic, following some perturbation its state will explore its entire Hilbert space, showing little dependence on the precise Hamiltonian parameters and washing out long-time correlation functions~\cite{Srednicki98Approach}.
This suggests that the criteria for a system to be learnable is that it correspond to the breakdown of ergodicity.
This breakdown has been well-studied in many-body physics for many years, most famously in the case of many-body localization~\cite{Basko06Metal,Abanin19Many}, and corresponds to many interesting phenomena such as area law entanglement scaling~\cite{Serbyn13Local,Bauer13Area} and the emergence of fractal many-body wavefunctions~\cite{Pino15Nonergodic,Altshuler16Nonergodic,Altshuler18Nonergodic,Faoro19Nonergodic,Mace19Multifractal}.
To the best of our knowledge, little has been done to tie these fundamental physics concepts to the notion of learnability of a quantum system.

When considering quantum Hamiltonian learning as a beyond-classical experiment, there are actually two classically-intractable quantum computations being performed.
The first is the Hamiltonian evolution itself (e.g. NMR experiment), which is an analog experiment being performed by the spectrometer and the sample.
This produces a set of data which is beyond-classical whenever the experiment is beyond-classical. 
(Beyond-classical does not necessarily require the experiment to be BQP-hard; many of the algorithms that we will consider in this work lie in the DQC-1 complexity theory class~\cite{Knill98Power,Somma08Parameter}.)
A key part of this work lies in identifying those NMR experiments that are classically challenging to simulate, which is one quality distinguishing our work from previous suggestions to study classically tractable NMR signals on a quantum device~\cite{Sels19Quantum}. The second classically-intractable quantum computation is to learn the quantum Hamiltonian from the experimental data, which can be executed on a digital quantum computer.
We assume that for the foreseeable future the connection between the spectrometer and the computer are classical.
This prevents the quantum Hamiltonian learning technique of Ref.~\cite{Wiebe14Quantum, Wiebe14Hamiltonian, Wang17Experimental}, or algorithms that require access to the quantum state~\cite{Huang21Power,Huang21Information} being implemented.
(Access to a quantum connection between spectrometer and computer would be of immense interest if it could be achieved, as it could provide additional exponential speedups~\cite{Huang21Information}.)
Alternatively, one can approach this problem by generating classically hard spectra (from a quantum simulation or NMR experiment) and using this to train a classical machine agent to infer Hamiltonians from spectra in the same phase (in a manner similar to Ref.~\cite{Huang21Provably}).
However, this requires access to this additional training data.

Given a sample and an NMR spectrometer,  chemists have many techniques at their disposal to infer molecular structure without requiring quantum computing assistance: magic angle spinning~\cite{Andrew58Nuclear,Lowe59Free}, dipolar decoupling and recoupling pulse schemes~\cite{Levitt86Composite,Bielecki89Frequency,Paravastu06Frequency,Tycko10Homonuclear}, polarization transfer methods~\cite{Ernst10Adiabatic}, and choices of different spin species and use of heteronuclear NMR couplings~\cite{Jaroniec10Dipolar}.
However, these methods throw away or approximate information about a system that is potentially valuable for characterization.
Here we show that quantum computers can provide an extra tool in the NMR toolbox, and in doing so open up a realm of novel NMR experiments that have not been available before.

In this paper we propose a protocol to learn the nuclear spin Hamiltonian of a molecular or material system using a digital quantum computer and time-resolved measurements from an NMR spectroscopy experiment.
We expect this protocol to present a beyond-classical quantum computation (in lieu of classical access to additional data such as a training set for a machine-learning algorithm) when the dataset from the NMR experiment is hard to classically simulate.
In section~\ref{sec:learning_problem}, we design quantum algorithms to estimate the cost function, Jacobian and Hessian of the learning problem.
We attempt to identify those systems and situations where a beyond-classical application can be found, following some general discussions of learnability in section~\ref{sec:learnability}.
In section~\ref{sec:FT_and_NISQ}, we describe and cost circuits to implement our quantum algorithms in both fault-tolerant and NISQ cost models.
We find that in both cost models the effect of integration error can be logarithmically suppressed or better, and that the ability to run deep coherent circuits in FT yields polynomial speedups in terms of various problem parameters; the duration and error in the experiment to be simulated, and the desired error in the final gradient itself.
In section~\ref{sec:NMR}, we identify NMR spectroscopy of proteins within cell walls or other membranes as one potential application, as the physical pinning of these systems within a membrane prevents the tumbling that would wash away strong correlations in solution.
In section~\ref{sec:ubiquitin}, we study an example protein, ubiquitin, as a benchmark with known molecular structure.
We identify sets of clusters of $^1$H spins within this molecule with strong intra-cluster coupling and weak coupling to the environment, that should produce a strongly-coupled signature able to be studied by a quantum device.
We calculate the multifractal dimension of small clusters to study their ergodic to non-ergodic phase transition as the dipolar term is suppressed (e.g. by magic angle spinning or decoupling pulse schemes), and find the `quantum-feasible' region to require a suppression factor between around $\alpha=5$ and $\alpha=100$ (assuming a background magnetic field of $23.5$~T, corresponding to a proton frequency of $1$~GHz).
This gives a large window within which quantum computers could be expected to assist in NMR interpretation.
We demonstrate the application of our learning algorithm to small clusters within this region, demonstrating its convergence on a small spin cluster in the presence of sampling noise.
Finally, we show a direct correspondence between the loss of ergodicity (as measured in the multifractal dimension of ubiquitin spin clusters as their dipolar coupling is suppressed) and the onset of learnability (as measured by the analytical Hessian of our learning problem at the global minimum).
To the best of our knowledge, this is the first demonstration of a clear connection between the notion of fractal eigenstates in a quantum system and its learnability by quantum or classical means.

\section{Quantum-assisted Hamiltonian learning}\label{sec:learning_problem}

We now consider the problem of learning a Hamiltonian $H$ of some system from a set of time-resolved experimental data $S_x(t)$, where, $x$ indexes different sets of experiments.
As the Hamiltonian of a system dictates the time dynamics, this is a natural thing to learn from time series data; we will discuss later how one can infer molecular structure from a Hamiltonian of nuclear spins.
Each experiment consists of an initial state preparation $\rho_x$, time evolution by $H$ plus an external time-dependent driving field $H_x(t)$, and final measurement of some observable $O_x$.
The signal $S_x(t)$ is then given by
\begin{equation}
    S_x(t) = \mathrm{Trace}[U_x(t,0)\rho_xU_x^{\dag}(t,0)O_x],\label{eq:signal}
\end{equation}
where $U_x(t_2,t_1)$ is the time evolution operator generated by the Hamiltonian $H+H_x(t)$ from $t=t_1$ to $t_2$
\begin{equation}
    U_x(t_2,t_1)=\mathfrak{T}\exp\left\{i\int_{t_1}^{t_2}\big[H+H_x(t)\big]dt\right\},
\end{equation}
where $\mathfrak{T}$ is the time-ordering operator.
Note that one may consider $\rho_x=\rho$ and $O_x=O$ independent of the experiment $x$ by encoding preparation and measurement terms onto the driving Hamiltonian $H_x(t)$.
(As we will discuss later, this is often an accurate description of a real-world NMR experiment.)
Alternatively, if the driving Hamiltonian $H_x(t)$ is only used for preparation and measurement, this may be encoded entirely in $\rho_x$ and $O_x$, setting $H_x(t)=0$ and $U_x(t_2,t_1)=U(t_2,t_1)=e^{iH(t_2-t_1)}$.

To define our learning problem, we write our system Hamiltonian in the form
\begin{equation}
    H=\sum_nh_nV_n,
\end{equation}
where $h_n$ are a set of parameters and $V_n$ a set of Hermitian operators.
We then consider the case where some or all of the $h_n$ are unknown, and we wish to estimate these by a set $\{\bar{h}_n\}$; this defines our learning problem.
Given some prior $\{h^{(0)}_n\}$ with standard deviation $w_n$, and assuming that each datapoint $S_x(t)$ is drawn from a normally-distributed experimental population with standard deviation $\sigma_{x,t}^2$, the maximum-likelihood estimation of the true parameters can be found by minimizing the cost function
\begin{equation}
    C\big[\bar{H}\big] = \sum_n\frac{(\bar{h}_n-h^{(0)}_n)^2}{2\omega_n^2}+\sum_{x,t}\frac{\big(\bar{S}_x(t)-S_x(t)\big)^2}{2\sigma_{x,t}^2}\label{eq:cost_function},
\end{equation}
where $\bar{S}_x(t)=\mathrm{Trace}[\bar{U}_x(t,0)\rho_x\bar{U}^{\dag}_x(t,0)O_x]$ is the estimated signal with our estimates of the parameters $\bar{h}_n$.
(Throughout this work, we use bars to denote quantities derived from estimated parameters rather than hidden ones.)
Though $\bar{S}_x(t)$ cannot be estimated on a classical device, implementing it on a quantum computer simply requires a circuit to simulate the time evolution $\bar{U}_x(t,0)$.
However, performing such an optimization gradient-free on a higher-dimensional surface is a costly endeavour.
The first key result of this work is to give a practical form for the gradient and Hessian of Eq.~\ref{eq:cost_function}
\begin{align}
    \frac{d C[H]}{d\bar{h}_n}&=\frac{1}{w_n^2}(\bar{h}_n-h^{(0)}_n)\nonumber\\
    &+\sum_{x,t}\frac{i}{\sigma^2_{x,t}}\big[\bar{S}_x(t)-S_x(t)\big]\bar{J}^n_{x}(t)\label{eq:cost_function_derivative}\\
    \frac{d^2 C[H]}{d\bar{h}_nd\bar{h}_m}&=\frac{\delta_{nm}}{w_n^2}-\sum_{x,t}\frac{1}{\sigma_{x,t}^2}\nonumber\\&\hspace{-0.3cm}\times\Big[\bar{J}^n_{x}(t)\bar{J}^m_{x}(t)+2\big[\bar{S}_x(t)-S_x(t)\big]\bar{K}^{n,m}_{x}(t)\Big],\label{eq:cost_function_second_derivative}
\end{align}
where we define
\begin{align}
    \bar{J}_x^n(t)&=\int_0^tds\,\bar{j}_x^n(t,s),\\
    \bar{j}_x^n(t,s)&=\mathrm{Trace}\Big[O_x\big[\bar{V}_{n,x}(t,s),\bar{\rho}_x(t)\big]\Big],\label{eq:Jxdef}\\
    \bar{K}_x^{n,m}(t)&=\int_0^tds\int_0^sdr\,k_x^{n,m}(t,s,r),\\
    \bar{k}_x^{n,m}(t,s,r)&\nonumber\\
    =\mathrm{Trace}&\bigg[O_x\Big[\bar{V}_{n,x}(t,s),\big[\bar{V}_{m,x}(t,r),\bar{\rho}_x(t)\big]\Big]\bigg],\label{eq:Kxdef}
\end{align}
and $\bar{V}_{n,x}(t,s)=\bar{U}_x(t,s)V_n\bar{U}^{\dag}_x(t,s)$ is the (estimated) operator $V_n$ evolved forwards in time from $s$ to $t$, and $\bar{\rho}_x(t)=\bar{U}_x(t,0)\rho_x\bar{U}_x^{\dag}(t,0)$ is the (estimated) state $\rho_x$ at time $t$.
As the term dependent on $\bar{K}_x^{n,m}(t)$ in Eq.~\ref{eq:cost_function_second_derivative} disappears in the limit $\bar{S}_x(t)\rightarrow S_x(t)$, it may be practical when near the global minimum of $C[H]$ to approximate
\begin{equation}
    \frac{d^2 C[H]}{d\bar{h}_ndh_m}=\frac{\delta_{nm}}{w_n^2}-\sum_{x,t}\frac{\bar{J}_x^n(t)\bar{J}_x^m(t)}{\sigma_{x,t}^2},\label{eq:cost_function_second_deriv_approx}
\end{equation}
which may be obtained at no extra cost to the gradient estimation (assuming $\bar{S}_x(t)$ and $\bar{J}_x^n(t)$ are measured to the same relative precision).
This is important, as we can approximate the covariance matrix $\Sigma$ of our final estimation of the $\{\bar{h}_n\}$ as 
\begin{equation}
\Sigma=[\nabla^2_hC]^{-1}.
\end{equation}
These equations may be alternatively derived via optimal control theory, which yields a conjugate field to the state $\bar{\rho}_x(t)$ that is generated by deviations $\bar{S}_x(t)-S_x(t)\neq 0$ and propagates backwards in time via the Schr\"{o}dinger equation.

\section{Quantum and classical learnability}\label{sec:learnability}

\begin{figure}[h!]
    \centering
    \includegraphics[width=\columnwidth]{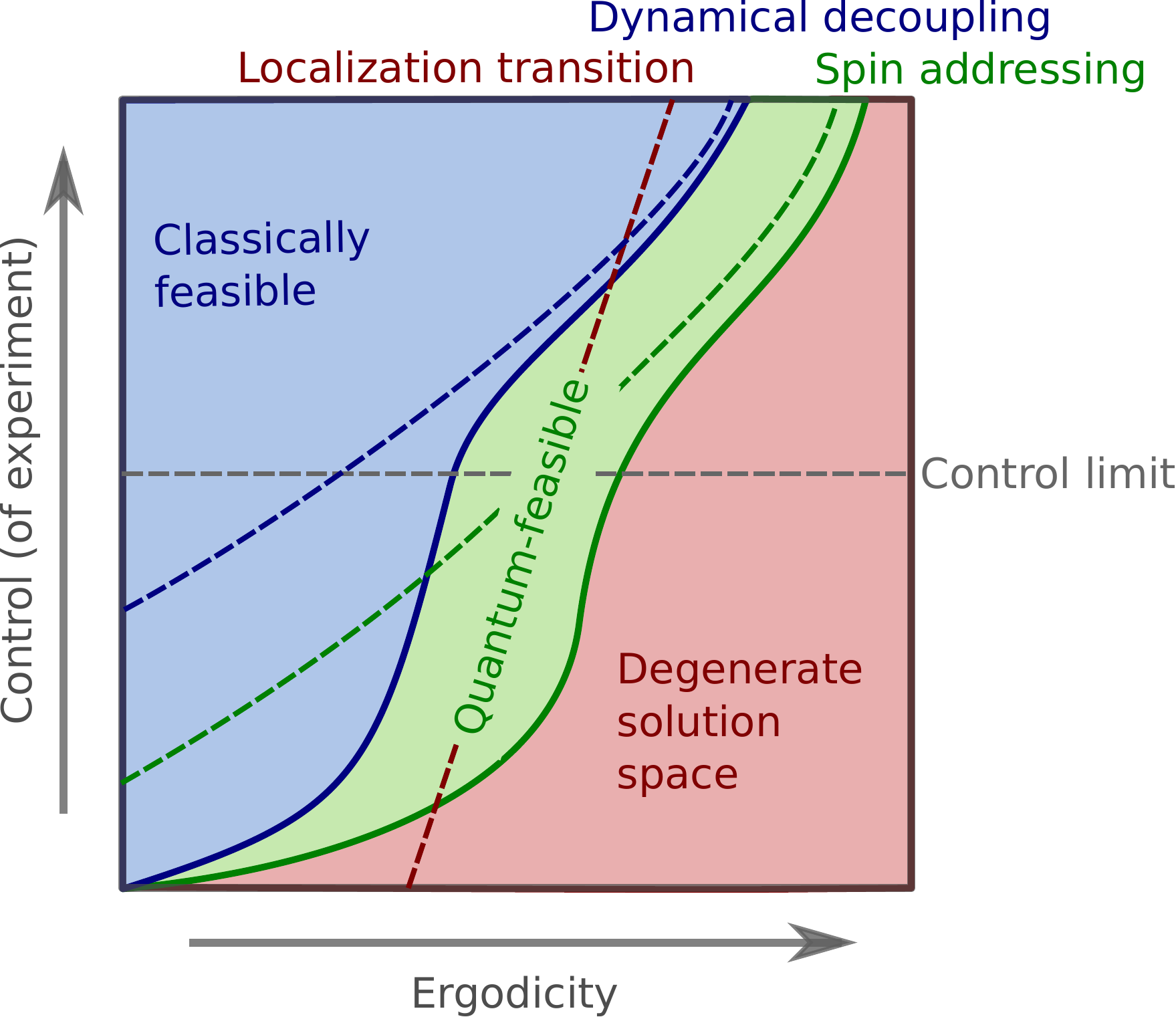}
    \caption{Cartoon of the ``phases of learnability'' of a quantum Hamiltonian. In the red region, the set of experimental data is insufficient to distinguish candidate Hamiltonians.
    This becomes increasingly likely for systems beyond a localization transition (red dashed line), as after this point experiments such as local correlators tend to provide little information about the system structure.
    In the blue region, the experimental data is sufficient to learn the system's structure, but the data processing may be achieved classically, rendering a quantum computer unnecessary.
    This classical processing may either be achieved by the system being well-approximated by a classically-computable model, or by the experiment having sufficient control (grey dashed line) to isolate smaller subsystems.
    At the limit of good control, an experiment has the ability to dynamically decouple individual terms (blue dashed line), after which techniques such as those in Refs.~\cite{Somma08Parameter,Wang15Hamiltonian} can be used to estimate terms individually.
    A simpler limit is to simply have the ability to address individual (or small frequency regions of) spins in the system (green dashed line), which renders the system more learnable.}
    \label{fig:learnability}
\end{figure}

In the following sections, we will describe how the procedure outlined in section~\ref{sec:learning_problem} may be developed into a complete quantum-assisted algorithm to learn Hamiltonian parameters via gradient optimization.
However, there are two issues that limit the usefulness of our proposal to implement these algorithms on a quantum computer.
We summarize these issues and give a sketch of the region in parameter space where the techniques we develop in this work are relevant in Fig.~\ref{fig:learnability}

The first issue is that the experimental data taken may not contain enough information to learn the desired couplings at all.
A large amount of nature is chaotic, or ergodic --- where the system tends to approximately explore its entire phase space.
In such ergodic systems, properties such as local correlation functions die off quickly without any dependence on the internal structure~\cite{Srednicki98Approach,Mi21Information}.
When this is the case the set of e.g. local spin-spin correlations generated by two different Hamiltonians may be indistinguishable up to corrections smaller than any experimental noise.
This makes learning impossible.
Given sufficient local disorder (relative to the strength of spin-spin interactions), systems tend to localize, yielding a non-ergodic many-body localized regime characterized by an absence of transport~\cite{Basko06Metal}.
This lack of transport prevents local correlators decaying, resulting in a time-resolved measurement that may be used to distinguish between, and thus learn, different Hamiltonians.
In certain systems, between the fully localized and ergodic regimes there has been proposed an intermediate non-ergodic regime, where the system explores a large fraction of its Hilbert space but does not completely thermalize~\cite{Pino15Nonergodic,Altshuler16Nonergodic,Altshuler18Nonergodic,Faoro19Nonergodic}.
In these situations, some learning should also be possible.
The transition to ergodicity and the loss of learnability may be mitigated somewhat given the ability to perform more complicated experiments (e.g. magic angle spinning~\cite{Andrew58Nuclear, Lowe59Free}, decoupling pulses~\cite{Tycko10Homonuclear}, or other composite pulse sequences~\cite{Jones09Composite}).
We summarize these notions going from left to right in Fig.~\ref{fig:learnability}.
Here, the x-axis denotes a rough measure of ergodicity of an arbitrary system (e.g. the localization length left of the localization transition, and the rate of entanglement growth on the right).
This can in principle be changed by experimental control (i.e. by suppressing the dipolar interaction strength), which suggests that the localization transition should not be a vertical line.

The second issue we face is that a quantum computer may not be required to interpret the spectrum of a given Hamiltonian: it may be entirely possible to solve the learning problem with a classical device~\cite{Elsayed15Effectiveness,Starkov18Hybrid}.
While this reduction in complexity is a good thing for the experiment in question; nevertheless it limits the utility of using quantum computers in that context.
The Hamiltonian learning problem may be solved classically for two reasons.
Firstly, the Hamiltonian itself may be classically simulatable, or approximately simulatable.
For example in a many-body localized system, the forward-scattering approximation or other perturbative expansions may be sufficiently accurate for learning.
This suggests that a quantum computer will find the most relevance studying either intermediate non-ergodic phases that are not completely localized, or the region in the proximity of a direct many-body localization transition where the localization length is too large for classical simulation.
The second reason why the Hamiltonian learning problem may be solved classically is if the experiment is controllable enough to isolate smaller subsystems or otherwise simplify the system.
This can for instance be achieved if one has the ability to spatially resolve individual spins with a magnetic field.
In this case it is possible to apply dynamic decoupling pulse sequences that isolate local Hamiltonian terms while cancelling out the remainder in an experiment~\cite{Somma08Parameter,Ajoy13Quantum,Wang15Hamiltonian}, making local characterization possible.
We summarize these notions along the y-axis of Fig.~\ref{fig:learnability}; as one gains more control over an experiment it becomes possible to learn Hamiltonians in more systems classically, until the barrier of dynamic decoupling is reached and all Hamiltonians are classically learnable.

These two concerns leave our quantum Hamiltonian learning algorithm with only a ``Goldilocks'' zone of applicability.
We are interested in those experiments where we have some control over our input state and Hamiltonian, but not those where enough control is available to isolate individual terms.
We are also interested in those experiments where our system is somewhat delocalized, but not completely.
Experimentally, this can be summarized by saying that we can study those systems where some signal can be extracted, but where that signal is complicated by features (e.g. spectral line shifts) that cannot be easily understood perturbatively.

\subsection{Robustness of learning}\label{sec:robust_learning}

Assuming access to the derivatives in Sec.~\ref{sec:learning_problem}, one may in principle solve the Hamiltonian learning problem using many well-known gradient-based or Hessian-based optimization techniques.
(Using the approximate Hessian in Eq.~\ref{eq:cost_function_second_deriv_approx} for minimization results in the well-known Levenberg-Marquardt algorithm~\cite{Levenberg44Method, Marquardt63Algorithm}.)
One may ask whether this can be made robust, to avoid being stuck in local minima.
Similar questions have been asked and answered previously for single parameter estimation~\cite{Somma08Parameter}, quantum phase estimation~\cite{Kimmel15Robust}, and device calibration~\cite{Neill21Accurately, Google20Observation}.
We give a sketch of an argument here for the robustness of our algorithm given a time-independent Hamiltonian ($H_x(t)=0$) that works under the assumption that we start with an initial guess $h_n^{(0)}$ sufficiently close to our true parameters $h_n$.
We stress that this is not a proof, and examining the landscape around the global minimum of our learning problem is a clear target for future work.
 
In this case, we may work in the eigenbasis $|\xi_a\rangle$ of the system Hamiltonian $H|\xi_a\rangle=E_a|\xi_a\rangle$.
Inserting two resolutions of the identity, our signal then takes the form $S_x(t)=\sum_{a,b}s^{a,b}_x(t)$, where
\begin{equation}
    s^{a,b}_x(t)=\langle\xi_a|\rho_x|\xi_b\rangle\langle\xi_b|O_x|\xi_a\rangle e^{i(E_a-E_b)t}.
\end{equation}
If our estimate deviates by some parameter $h_n\rightarrow\bar{h}_n=h_n+\delta$, then to lowest order in perturbation theory our estimated signal takes the form
\begin{align}
    \bar{S}_x(t)&=\sum_{a,b}\bar{s}^{a,b}_x(t)+\delta X,\\
    \bar{s}^{a,b}_x(t)&=s^{a,b}_x(t)e^{it\delta(\langle\xi_a|V_n|\xi_a\rangle - \langle\xi_b|V_n|\xi_b\rangle)}\\
    X&=\sum_{a,b,c, a\neq c}\frac{e^{-iE_bt}(e^{iE_at}-e^{iE_ct})}{E_a-E_c}\nonumber\\
    &\hspace{1.5cm}\times\langle\xi_{c}|\rho_x|\xi_b\rangle\langle\xi_b|O_x|\xi_a\rangle\langle\xi_a|V_n|\xi_c\rangle\nonumber\\
    &+\sum_{a,b,c, b\neq c}\frac{e^{iE_at}(e^{-iE_bt}-e^{-iE_ct})}{E_b-E_{c}}\nonumber\\
    &\hspace{1.5cm}\times\langle\xi_{a}|\rho_x|\xi_c\rangle\langle\xi_c|V_n|\xi_b\rangle\langle\xi_b|O_x|\xi_a\rangle.
\end{align}
Note here that $X$ is independent of $\delta$.
The second term in our cost function (Eq.~\ref{eq:cost_function}) takes the form
\begin{align}
    &\sum_{x,t}\frac{\big(\bar{S}_x(t)-S_x(t)\big)^2}{2\sigma_{x,t}^2}=\sum_{x,t}\frac{1}{\sigma_{x,t}^2}\bigg[\delta X\nonumber\\
    &+\sum_{a,b}s_x^{(a,b)}(t)\left(1-e^{it\delta(\langle\xi_a|V_n|\xi_a\rangle - \langle\xi_b|V_n|\xi_b\rangle)}\right)\bigg]^2 \, .
\end{align}
This oscillates as a function of $\delta$ with a frequency bounded by $4t\max\langle\xi_a|V_n|\xi_a\rangle$, which is independent of the system size.
This implies that we know that local minima in our parameter space must be separated by at least $[4t\max\langle\xi_a|V_n|\xi_a\rangle]^{-1}\leq [4t\max_n\|V_n\|]^{-1}$.
Flipping this around, let us suppose we know our initial guess $h_n^{(0)}$ of our parameters $h_n$ lies within some $\delta\leq \sum_n|h_n-h_n^{(0)}|$,
we can perform robust Hamiltonian learning by first learning $H$ from only experiments at times
\begin{equation}
    t < t_{\max} = \frac{\pi}{4\delta\max_n\|V_n\|}.
\end{equation}
After converging on this data we may estimate the variance of our parameter guess (using Eq.~\ref{eq:cost_function_second_derivative}), refine our estimate of $\delta$, and increase the range of allowed $t$.
Assuming that estimation at each $t_{\max}$ yields an error $\delta_{\mathrm{new}}\leq c\pi/(4\max_n\|V_n\|t_{\max})$ for some $c<1$, repeating this procedure over multiple orders will converge to some final error $\epsilon$ in $\mathcal{O}(\log(1/\epsilon))$ iterations of this procedure. 
If our initial uncertainty $\sum_n|h_n-h_n^{(0)}|$ grows with the system size (which can simply be due to the increase in the number of parameters), this will necessarily shrink the initial choice of $t_{\max}$ by the same amount.
However, for a local Hamiltonian we expect the number of parameters and the magnitude of terms to only grow linearly in $N$.

The ability to estimate out to large times allows us to avoid learnability issues in localized systems where small long-range couplings contribute mostly to the off-diagonal part of the Hamiltonian (and thus do not affect the eigenstructure significantly).
However, it is of no help in ergodic systems where signals $S_x(t)$ disappear quickly with $t$.
As we will see, in those (ergodic) systems learning anything would be a significant challenge.

\section{Quantum algorithms}\label{sec:FT_and_NISQ}

In this section, we outline algorithms to efficiently calculate our cost function $C[H]$ and its first and second derivatives using a quantum computer.
Quantum algorithm optimization is significantly different when targeting noisy near-term vs fault-tolerant long-term devices; we will present algorithms for both situations.
In either case, we will discretize the integral in Eq.~\ref{eq:Jxdef}
\begin{equation}
    \int_0^tdsf(s)\sim \sum_{i=0}^{I-1}z_if\left(s_i\right),\label{eq:numerical_integration}
\end{equation}
where the weights $z_i>0$ are chosen such that $\sum_iz_i=t$.
(A $2$-dimensional discretization is similarly possible for the integral $K_x^{n,m}(t)$.)
There are many possible methods for choosing both the weights and the points $s_i$; one may use a simple trapezoidal or midpoint rule, or more complicated Gaussian quadrature methods, or one may take a Monte Carlo approach and choose the points at random.
Each method incurs a discretization error that goes to $0$ as $I\rightarrow\infty$.
For both the FT and NISQ quantum methods that we propose, cost depends primarily on $[\sum_iz_i]=t$, and has only at most a logarithmic dependence on $I$.
This allows us to choose our method of integration for simplicity or ease in circuit design.

\subsection{Algorithms for near-term quantum computers}\label{sec:algorithms}

\begin{figure}
    \centering
    \includegraphics[width=\columnwidth]{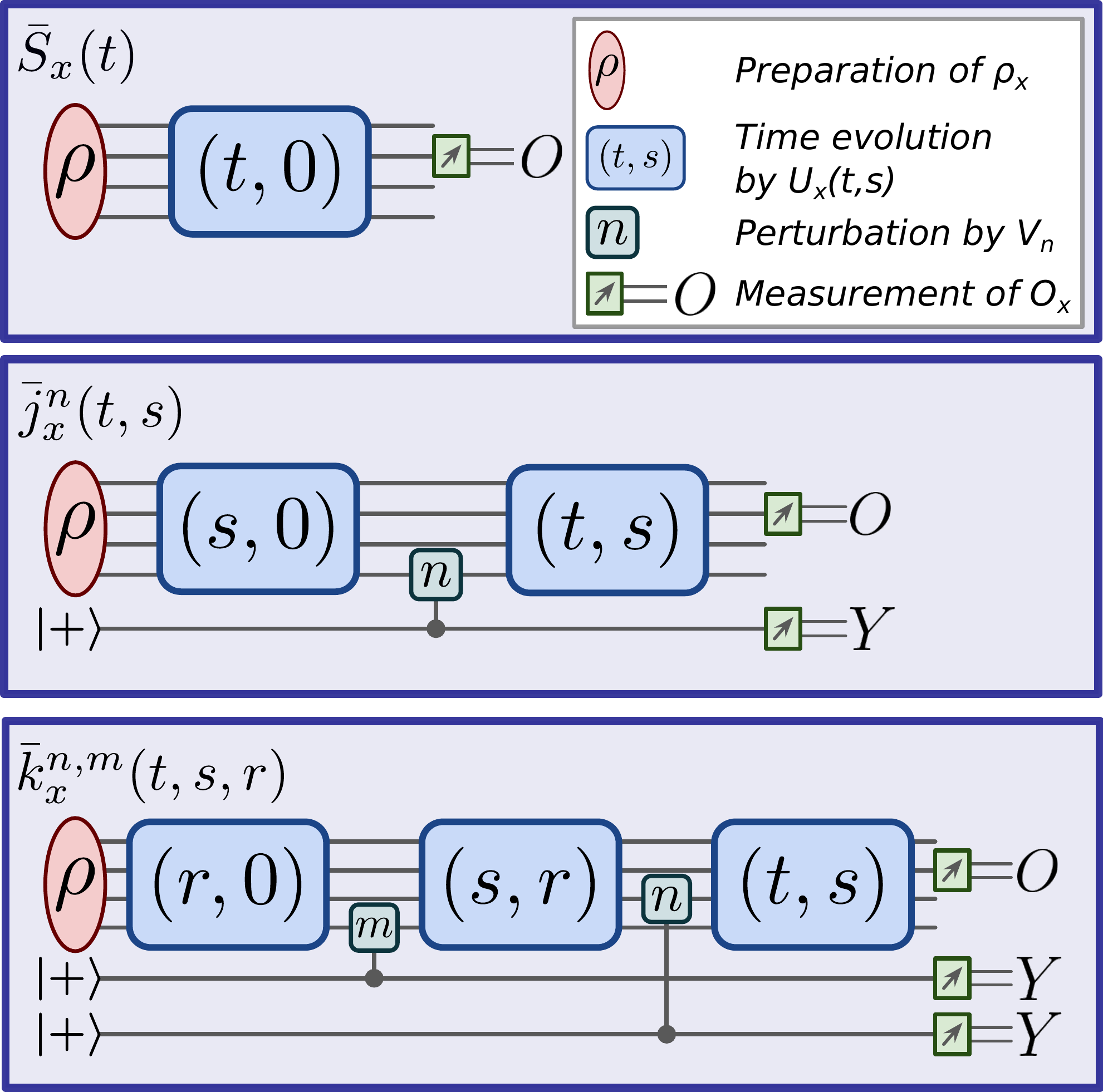}
    \caption{Circuits to estimate the signal $\bar{S}_x(t)$ (top) and the integrands $\bar{j}_x^n(t,s)$ (middle) and $\bar{k}_x^{n,m}(t,s,r)$ (bottom) that are required to calculate the first and second derivatives of our cost function $C[H]$ (Eq.~\ref{eq:cost_function}).
    For ease of viewing, we suppress labels in the circuits themselves (see legend for clarification).
    Circuits assume access to a preparation of $\rho$, and a means to simulate $U_x(t,s)$ (without control) and to implement controlled perturbations $V_n$.
    The desired integrand can be found to be the expectation value of the product of the indicated operators, which (in NISQ) must be read out by repeated preparation and measurement.}
    \label{fig:NISQ_circuits}
\end{figure}

In the current NISQ era we want to run the shortest quantum circuits possible for any application.
To this end, we propose estimating the signal $\bar{S}_x(t)$, and the integrands $\bar{j}^n_x(t)$ and $\bar{k}^{n,m}_x(t)$ on a quantum computer, and performing the integration and summation to yield $\frac{dC[H]}{dh_n}$ and $\frac{d^2C[H]}{dh_ndh_m}$ classically.
The signal $\bar{S}_x(t)$ is already in the form where it can be read as the expectation value of a quantum state following the application of a unitary circuit (Fig.~\ref{fig:NISQ_circuits}, top).
This requires that it take the form $\mathrm{Trace}[\mathcal{U}\Xi\mathcal{U}^{\dag}\mathcal{M}]$ for a quantum state $\Xi$, unitary $\mathcal{U}$, and hermitian operator $\mathcal{M}$.
(This can be compared directly to Eq.~\ref{eq:signal}.)
The integrands $\bar{j}^n_x(t)$ (Eq.~\ref{eq:Jxdef}) and $\bar{k}^{n,m}_x(t)$ (Eq.~\ref{eq:Kxdef}) are not quite of this form.
However, the integrands may be put in the correct form by adding control qubits to enlarge the Hilbert space, and then using the so-called generalized Hadamard test; this results in the middle and bottom circuits of Fig.~\ref{fig:NISQ_circuits}.
This circuit construction uses the identity
\begin{align}
    &i\mathrm{Trace}[O[U,\rho]] \nonumber\\&= 2\mathrm{Trace}\Big[(\mathrm{c-}U)(|+\rangle\langle +|\otimes\rho)(\mathrm{c-}U^{\dag})(Y\otimes O)\Big],
\end{align}
to transform a commutator into the above form (with $Y\otimes O=\mathcal{M}$, $\mathrm{c-}U=\mathcal{U}$ and $|+\rangle\langle +|\otimes\rho = \Xi$).
Here the symbol $\mathrm{c-}U := I\oplus U$ denotes  the unitary $U$ controlled by the control qubit.
These circuits require only local control of the $V_n$ unitary (see below) and uncontrolled time evolution, making them rather NISQ friendly (using e.g. the randomized Trotterization methods of Ref.~\cite{Campbell19Random} to simulate the time evolution).
We assume in Fig.~\ref{fig:NISQ_circuits} that the terms $V_n$ are unitary (i.e., tensor products of Pauli operators for pairs of spins), but if this is not the case one may write $V_n$ as a linear combination of unitary operators, execute the circuits for each unitary component separately, and sum the resulting expectation values to yield the desired result~\cite{Faehrmann21Randomizing}.

The circuits in Fig.~\ref{fig:NISQ_circuits} assume the ability to prepare the initial states $\rho_x$.
In the applications we consider in this paper, these will be mixed diagonal states in the computational basis.
The measurement operators $O_x$ will similarly be diagonal in the computational basis.
Preparing mixed states requires that we average over many pure state preparations.
Consider the case where we prepare a computational basis state $|n\rangle$, then perform a circuit $U$ and measure the expectation values of a set of $O_x$ in parallel, yielding a set of estimates of $\langle n|U^{\dag}O_xU|n\rangle$.
If we have repeated this independently for all computational basis states, and each $\rho_x$ is diagonal in the computational basis, we can write
\begin{align}
    \mathrm{Trace}[U\rho_xU^{\dag}O_x]&=\sum_{m,n}\langle n|\rho_x|m\rangle\langle m|U^{\dag}O_xU|n\rangle\nonumber\\
    &=\sum_{n}\langle n|\rho_x|n\rangle\langle n|U^{\dag}O_xU|n\rangle.
\end{align}
As we know the initial distributions $\langle n|\rho_x|n\rangle$, the estimations of $\langle n|U^{\dag}O_xU|n\rangle$ may be used to compute the target trace.
In practice we do not need to prepare all states; it suffices to sample from a distribution proportional to $|\langle n|\rho_x|n\rangle|$.
In state-of-the-art quantum experiments this presents a small difficulty, as uploading a new pulse sequence to re-prepare each state may be impractical.
One solution may be to initially prepare qubits in the $|+\rangle$ basis and measure them prior to performing a simulation, which results in a new preparation each time.

Repeating the above procedure at multiple points $s_i$ and $r_i$ allows for parallel estimation of $\bar{J}_x^n(t)$ or $\bar{K}_x^{n,m}(t)$ via Eq.~\ref{eq:numerical_integration} (for fixed $n$ and $m$).
In principle in NISQ we are free to draw these times at random from the range $[0,t]$ and $[0,s]$ and choose the initial state.
(In practice changing evolution times is more difficult than choosing initial states; making this scheme more practical is an important direction for future work.)
In this case, each choice of starting state and time produces an independent random variable and Hoeffding's inequality may be applied.
For the estimation of $\bar{J}_x^n(t)$, $M$ repetitions of our experiment yields an estimator $\overline{\bar{J}_x^n(t)}$ that satisfies
\begin{equation}
    P\left(\left|\bar{J}_x^n(t)-\overline{\bar{J}_x^n(t)}\right|>\epsilon\right)\leq 2e^{-\frac{M\epsilon^2}{2\|O_x\|^2t^2}},\label{eq:NISQ_Hoeffding_bound}
\end{equation}
and the number of samples required to estimate this at a constant failure rate scales as $M=\widetilde{\mathcal{O}}(\epsilon^{-2}\|O_x\|^2t^2)$ (where we use $\widetilde{\mathcal{O}}$ to denote asymptotic complexity suppressing polylogarithmic factors).
Similarly, for the estimation of $\bar{K}_x^{n,m}(t)$, $M$ repetitions of our experiment yields an estimator $\overline{\bar{K}_x^{n,m}(t)}$ that satisfies
\begin{equation}
    P\left(\left|\bar{K}_x^{n,m}(t)-\overline{\bar{K}_x^{n,m}(t)}\right|>\epsilon\right)\leq 2e^{-\frac{M\epsilon^2}{\|O_x\|^2t^4}},
\end{equation}
and the number of samples required to estimate this at a constant failure rate scales as $M=\widetilde{\mathcal{O}}(\epsilon^{-2}\|O_x\|^2t^4)$.
This scaling in $t$ is to be expected, as the integrals tend to scale as $J_x^n(t)\sim t$, $K_x^{n,m}(t)\sim t^2$.
(To see this, note that the diagonal terms of $\int_0^tds\bar{V}_{n,x}(t,s)$ in the Hamiltonian basis grow linearly in $t$~\cite{Facchi03Unification,Tran21Symmetry}.)
This implies that we may estimate $\bar{J}_x^n(t)$ and $\bar{K}_x^{n,m}(t)$ to constant relative error with a number of samples independent of $t$.

We now give an analysis of the complexity of this sampling approach where we assume quantum simulation is performed under a Hamiltonian query model.
The no-fast-forward theorem~\cite{Berry07Efficient} requires a number of queries of $\Omega(t)$ to execute each of the circuits in Fig.~\ref{fig:NISQ_circuits}, so the total gate count to estimate $\bar{J}_x^n(t)$ and $\bar{K}_x^{n,m}(t)$ to error $\epsilon$ using these methods scales at best as $\widetilde{\mathcal{O}}(\epsilon^{-2}\|O_x\|^2t^3)$ and $\widetilde{\mathcal{O}}(\epsilon^{-2}\|O_x\|^2t^5)$ respectively.
By comparison, the total query count to estimate $\bar{S}_x(t)$ to error $\epsilon$ using the circuit in Fig.~\ref{fig:NISQ_circuits} is bounded asymptotically as $\widetilde{\mathcal{O}}(\epsilon^{-2}\|O_x\|^2t)$.
These estimates need to be combined to estimate the derivative in Eq.~\ref{eq:cost_function_derivative}.
To compute the total cost to estimate this to constant error, let us assume that $\bar{J}_x^n(t)\sim t$, $\bar{S}_x(t)\sim 1$, that $\sigma_{x,t}\sim\sigma$ and $\|O_x\|\sim 1$ are independent of $x$ and $t$, and that we can optimize the number of repetitions of each experiment to minimize the total query count.
Let us also assume that each experiment involves preparation and measurement in a commuting basis, and let us assume no covariance between parallel measurements.
Then, we find the total number of queries required to estimate single derivative terms to error $\epsilon$ is bounded asymptotically by (see App.~\ref{app:msmt_optimization} for details)
\begin{equation}
    \widetilde{\mathcal{O}}\left(\sigma^{-4}\epsilon^{-2}N_x\left[\sum_{\mathrm{sampled}\; t}t^{3/2}\right]^2\right),\label{eq:NISQ_scaling_general}
\end{equation}
where $N_x$ is the number of distinct experiments performed.
The evaluation of the sum over $t$ depends on whether $S_x(t)$ are sampled logarithmically sparsely (in which case $\sum_t t^{3/2}\sim T^{3/2}\log(T)$), or densely (in which case $\sum_t t^{3/2}\sim T^{5/2}$), where $T=\max(t)$.
In the former case, the total number of queries is bounded asymptotically by $\widetilde{\mathcal{O}}\left(\sigma^{-4}\epsilon^{-2}N_xT^3\right)$, whilst in the latter it is bounded asymptotically by $\widetilde{\mathcal{O}}\left(\sigma^{-4}\epsilon^{-2}N_xT^5\right)$.

\subsection{Improved estimation of the gradient term on a fault-tolerant quantum computer}\label{sec:ft}

In a fault-tolerant cost model, it is preferable to perform the integration, multiplication and summation over $t$ and $x$ in the second term of Eq.~\ref{eq:cost_function_derivative} entirely coherently.
This because on an error-corrected quantum computer the key resource to minimize is the total number of gates (i.e. the sum of the gate count of each circuit applied) rather than just the depth of the longest circuit.
We now outline how this may be achieved for the gradient term.
Assuming that $\rho_x=W_x|0\rangle\langle 0|W_x^{\dag}$, and using the fact that $\mathrm{Trace}(A)\cdot\mathrm{Trace}(B)=\mathrm{Trace}(A\otimes B)$, we have
\begin{widetext}
\begin{align}
    &\sum_{x,t}\frac{i}{\sigma^2_{x,t}}\bar{S}_x(t)\bar{J}_x^n(t)\sim\Bigg\langle0\otimes0\otimes +\Bigg|\sum_{x,t,l}\frac{z_{l,t}}{\sigma_{x,t}^2}\mathcal{U}^0(x,t,s_{l,t})\Bigg|0\otimes0\otimes +\Bigg\rangle\\
    &\mathcal{U}^0(x,t,s)=W^{\dag}_xU_x^{\dag}(t,0)O_xU_x(t,0)W_x\nonumber\\&\hspace{2cm}\otimes W_x^{\dag}U_x^{\dag}(s,0)\Big[|0\rangle\langle 0| - |1\rangle\langle 1|\otimes V_n\Big] U_x^{\dag}(t,s)O_xU_x(t,s)\Big[|0\rangle\langle 0|\otimes V_n + |1\rangle\langle 1|\Big]U(s,0)W_x,~\label{eq:LCU_decomposition1}
\end{align}
\end{widetext}
where the approximation is the approximation from our numerical integration.
One can confirm that $\mathcal{U}^0(x,t,s)$ is unitary as long as $V_n$ and $O_x$ are unitaries (and if this is not the case, they may be decomposed as a linear combination of unitaries themselves).
The second part of the second term in Eq.~\ref{eq:cost_function_derivative} requires multiplying by the experimental signal $S_x(t)$.
This signal then needs to be loaded onto the device; if done naively this could easily become the dominant cost in our circuit.
To lower this cost, we make the reasonable assumption that $S_x(t)$ consists of a small number $N_{\omega}\ll T$ of Fourier components
\begin{equation}
    S_x(t)=\sum_{k=1}^{N_{\omega}}a_{x,k}\cos(t\omega_{x,k} + \phi_{x}),
\end{equation}
where $\phi_x=0$ or $\phi_x=\pi/2$ and $a_{x,k}>0$ is expected from the $t=0$ behaviour of our signal.
Then, writing $\cos(t\omega_{x,k}+\phi_{x})=\frac{1}{2}(e^{i(t\omega_{x,k}+\phi_{x})}+e^{-i(t\omega_{x,k}+\phi_{x})})$, we have
\begin{widetext}
\begin{align}
    \sum_{x,t}\frac{i}{\sigma^2_{x,t}}&S_x(t)\bar{J}_x^n(t)\sim\Bigg\langle0\otimes +_0\otimes +_1\Bigg|\sum_{x,t,l,k}\frac{z_{l,t}a_{x,k}}{\sigma_{x,t}^2}\mathcal{U}^1(x,t,s_{l,t},k)\Bigg|0\otimes +_0\otimes +_1\Bigg\rangle\\
    \mathcal{U}^1&(x,t,s,k)=e^{-iZ_1(t\omega_{x,k}+\phi_x)}\nonumber\\
    &\times W_x^{\dag}U_x^{\dag}(s,0)\Big[|0_0\rangle\langle 0_0| - |1_0\rangle\langle 1_0|\otimes V_n\Big] U_x^{\dag}(t,s)O_xU_x(t,s)\Big[|0_0\rangle\langle 0_0|\otimes V_n + |1_0\rangle\langle 1_0|\Big]U_x(s,0)W_x,~\label{eq:LCU_decomposition2}
\end{align}
\end{widetext}
where we have labeled the operations acting on the different control qubits $0$ and $1$. Under the above assumptions, $\mathcal{U}^1(x,t,s)$ is also a unitary operator.
In the above, the $s_{l,t}$ and $z_{l,t}$ points are our integration points and weights respectively (following Sec.~\ref{sec:algorithms}), but allowing for the fact that the limits of integration (and thus both the points we should sample over and the total width we need to multiply by) are dependent on $t$.
Both summations may be then block encoded using standard LCU techniques~\cite{Childs12Hamiltonian}.
These require control registers $|x\rangle$,$|t\rangle$, $|l\rangle$, and $|k\rangle$ to encode the summation variables (as well as some additional registers we will introduce later), and SELECT and PREPARE unitaries.
(We assume here that our times $|t\rangle$ have some finite binary representation.)
These registers contain in turn $n_x\sim\log(N_x), n_t\sim\log(KT)$, $n_l\sim\log(L)$ and $n_k\sim\log(N_\omega)$ qubits, where $1/K$ is the precision to which we store our times $t$; see \append{error} for a detailed analysis of the truncation and discretization error.

\begin{figure}
    \centering
    \includegraphics[width=\columnwidth]{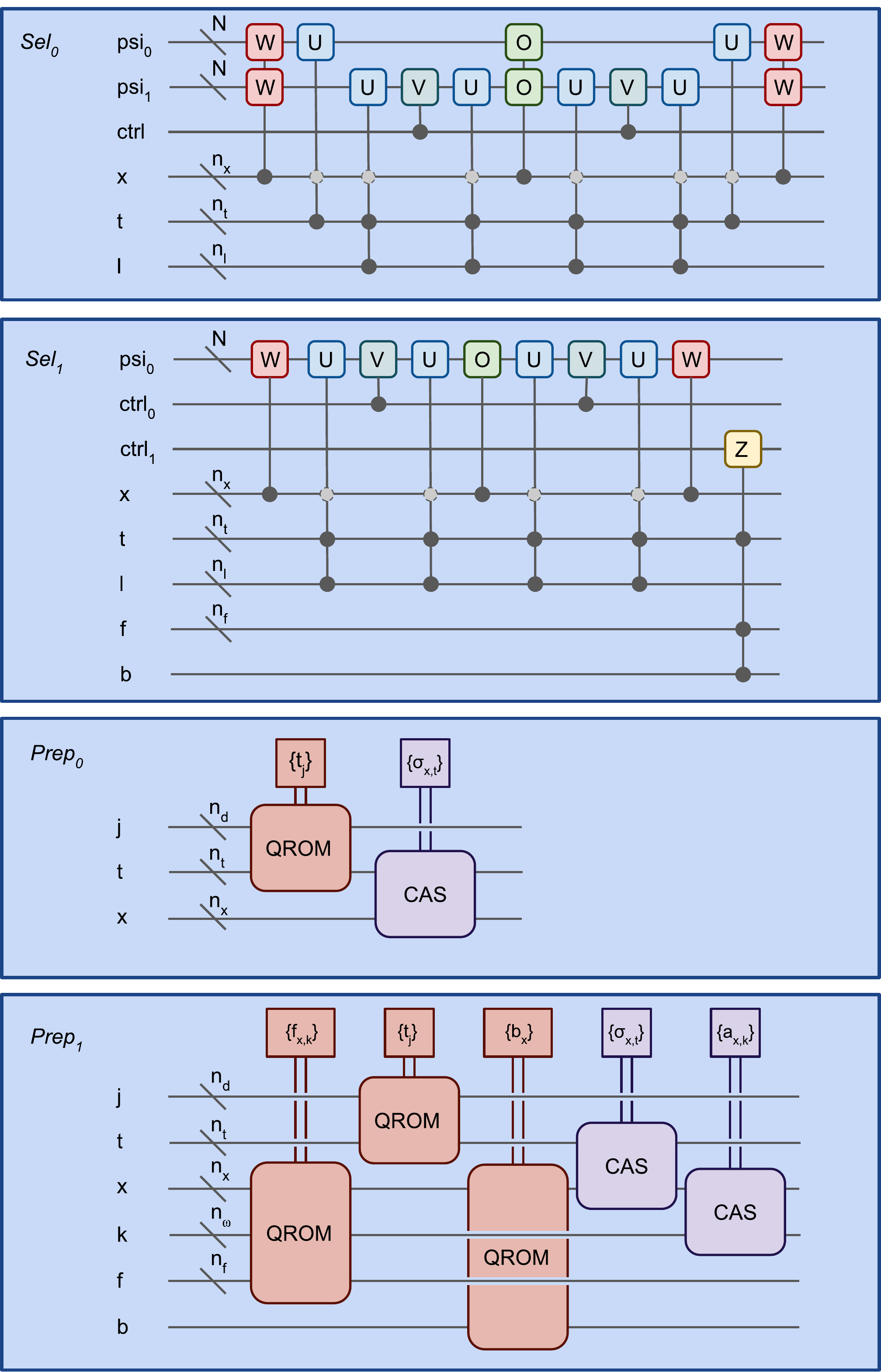}
    \caption{
    Circuit diagrams of the fault-tolerant oracles SEL$_0$, SEL$_1$, PREP$_0$ and PREP$_1$ described in this text. See text for details. Black circles on multi-qubit registers denote complex control procedures. Square boxes denote classical input to the system via QROM and coherent alias sampling (CAS)~\cite{Babbush18Encoding}. Subscripts are omitted from gates for ease of reading. The dashed circles on the control for $U$ in the SEL$_a$ circuits indicates control that is only needed if the time evolution during an experiment changes between experiments.}
    \label{fig:FT_circuits}
\end{figure}

The SEL$_0$ unitary selects the correct $\mathcal{U}^0(x,t,s_{l,t})$ unitary to implement based on the control register; in other words, $$\mathrm{SEL}_0=\sum_{x,t,l}|x\rangle|t\rangle|l\rangle\langle l|\langle t|\langle x|\mathcal{U}^0(x,t,s_{l,t})$$.
Similarly, $$\mathrm{SEL}_1=\sum_{x,t,l,k}|x\rangle|t\rangle|l\rangle|k\rangle\langle k|\langle l|\langle t|\langle x|\mathcal{U}^1(x,t,s_{l,t},k)$$.
In Fig.~\ref{fig:FT_circuits}, we show how this can be implemented using oracular access to $U_x(t,s)$, $O_x$, and $W_x$ (which we will shortly give implementations for).
The PREP$_a$ unitaries prepare the corresponding control states
\begin{align}
    |\Psi_c^0\rangle &=\frac{1}{\sqrt{\lambda_0}}\sum_{x,t,l}\sqrt{\frac{z_{l,t}}{\sigma_{x,t}^2}}|x\rangle|t\rangle|l\rangle,\label{eq:control_state1}\\
    \lambda_0 &= \sum_{x,t,l}\left|\frac{z_{l,t}}{\sigma_{x,t}^2}\right|,\label{eq:lambda0}\\
    |\Psi_c^1\rangle &=\frac{1}{\sqrt{\lambda_1}}\sum_{x,t,l,k}\sqrt{\frac{z_{l,t}a_{x,k}}{\sigma_{x,t}^2}}|x\rangle|t\rangle|l\rangle|k\rangle,\label{eq:control_state2}\\
    \lambda_1 &= \sum_{x,t,l,k}\left|\frac{z_{l,t}a_{x,k}}{\sigma_{x,t}^2}\right|,\label{eq:lambda1}
\end{align}
from an initial state $|0\rangle$ on the control register.
(Note that the absolute value in Eq.~\ref{eq:lambda0} and Eq.~\ref{eq:lambda1} are technically unnecessary as all summands are positive.)
We omit garbage registers in these steps for simplicity.
Given these, one can check that
{\small{
\begin{align}
    \langle 0|\mathrm{PREP}_0^{\dag}\;\mathrm{SEL}_0\;\mathrm{PREP}_0|0\rangle&\sim \frac{1}{\lambda_0}\sum_{x,t}\frac{i}{\sigma^2_{x,t}}\bar{S}_x(t)\bar{J}_x^n(t)\\
    \langle 0|\mathrm{PREP}_1^{\dag}\;\mathrm{SEL}_1\;\mathrm{PREP}_1|0\rangle&\sim \frac{1}{\lambda_1}\sum_{x,t}\frac{i}{\sigma^2_{x,t}}S_x(t)\bar{J}_x^n(t),
\end{align}}}\\
where the circuits act on the combined system and control register set.
For $a=0,1$, we may then use the overlap estimation algorithm of Ref.~\cite{Knill07Optimal} to estimate these values to error $\epsilon_a$ with confidence $1-\delta$ using $\mathcal{O}(\log(\delta^{-1})\epsilon_a^{-1})$ queries to $\mathrm{PREP}_a$, $\mathrm{SEL}_a$;
see \append{qoe} for a review of this algorithm.
The number of additional gates used in \cite{Knill07Optimal} are negligible compared to the cost of block encoding.
Setting $\epsilon_a={\cal O}(\epsilon/\lambda_a)$ allows us to obtain an estimate of $\frac{dC[H]}{dh_n}$ that is within $\epsilon$ with confidence $1-\delta$.
$\sum_iz_{l,t}=t$ when we have sampled at time $t$, so if we assume that $|S_x(t)|\leq 1$ and $\sigma_{x,t}=\sigma$, we have $\lambda_1=\lambda_2=\frac{N_x}{\sigma^2}\sum_{\mathrm{sampled}\;t}t$.
The number of oracle calls to PREPARE and SELECT then scales as (in comparison to Eq.~\ref{eq:NISQ_scaling_general})
\begin{equation}
    \widetilde{\mathcal{O}}\left(\sigma^{-2}\epsilon^{-1}N_x\sum_{\mathrm{sampled}\; t}t\right).\label{eq:FT_scaling_general}
\end{equation}
To make a heuristic comparison to the NISQ results, we again consider an oracular model.
Our SELECT oracles require time evolution by up to $T=\max(t)$, so a quantum simulation algorithm with linear scaling in the evolution time would make $\mathcal{O}(T)$ queries to the Hamiltonian oracle.
Thus, in the sparse sampling case the total number of oracle calls is bounded by $\mathcal{O}(\sigma^{-2}\epsilon^{-1}N_xT^2)$ (a saving of $\sigma^{-2}\epsilon^{-1}T$), while in the dense sampling case the total number of oracle calls is $\mathcal{O}(\sigma^{-2}\epsilon^{-1}N_xT^3)$ (a saving of $\sigma^{-2}\epsilon^{-1}T^2$).
We expect much larger savings for simulating concrete Hamiltonians using fault-tolerant quantum algorithms, as NISQ approaches \cite{Campbell19Random,Faehrmann21Randomizing} typically cannot achieve linear scaling in the simulation time and also have worse scaling in the target precision.

The complexity of the SELECT unitaries is dictated by the need to implement the controlled time evolution $\sum_{x,t,l}|x\rangle|t\rangle|l\rangle\langle l|\langle t|\langle x|U_x(s_{l,t},0)$ and $\sum_{x,t,l}|x\rangle|t\rangle|l\rangle\langle l|\langle t|\langle x|U_x(t,s_{l,t})$.
(We will discuss the initial preparation $\sum_x|x\rangle\langle x|W_x$ later.)
There are a wide range of fast quantum algorithms for simulating time evolution \cite{Berry15Taylor,Low2019hamiltonian,Low18Interaction,haah2021quantum,Berry07Efficient}; which choice is optimal depends on the details of the Hamiltonian being studied.
Here, we give an example implementation of the SELECT unitaries using higher-order product formulas, following the analyses developed in Ref.~\cite{Childs21Theory}.
We expect higher-order formulas to provide the fastest approach for simulating many spin Hamiltonians \cite{Childs18Toward,Childs21Theory}, at least asymptotically.
We assume for practical purposes here that our time evolution is experiment-independent --- $U_x(t,s)=U(t,s)=e^{iH(t-s)}$.
This removes the need to consider the $|x\rangle$ register in our implementation of SELECT.
Our implementation requires that we fix the discretization of the integral in Eq.~\ref{eq:Jxdef}.
We choose $L$ points $s_{l,t}=\frac{tl}{L}$, with even weights $w_l=\frac{t}{L}$.
The error in this approximation can be shown to be bounded by $O\left(\frac{t^2}{L}\|O_x\|\|[H,V_n]\|\right)$, which is negligible if $L\gg t^2\|O_x\|\|[H,V_n]\|$; see \append{error} for details.
The controlled time evolution part of the SELECT unitary then takes the form
\begin{align}
    \label{eq:controlled_evolution1}
    \mathrm{c-}U(s,0) &= \sum_{l=0}^L\sum_{t=0}^T|l\rangle|t\rangle\langle t|\langle l|e^{\frac{-ilt}{L}H},\\
    \label{eq:controlled_evolution2}
    \mathrm{c-}U(t,s) &= \sum_{l=0}^L\sum_{t=0}^T|l\rangle|t\rangle\langle t|\langle l|e^{\frac{-i(L-l)t}{L}H}.
\end{align}
We implement this by dividing the total time interval $[0,t]$ into $R$ Trotter steps and implementing a higher-order product formula in each step. To ensure that the simulation has error at most $\eta$, we take~\cite{Childs21Theory}
\begin{equation}
\label{eq:def_Trotter_number}
    R=X_1T(X_2T/\eta)^{o(1)},
\end{equation}
where the lower-case $o(1)$ here represents a constant that can be taken to be arbitrarily small, and the $X_1$ and $X_2$ coefficients depend on the system size and graph connectivity.
For a linear chain of $N$ qubits, we have $X_1\sim 1$ and $X_2\sim N$ \cite{Childs19Lattice}.
By contrast, assuming a model of clustered Hamiltonians~\cite{Peng20Simulating}
\begin{equation}
\label{eq:def_cluster}
    H=\sum_{\mathcal{K}}\sum_{k,k'\in\mathcal{K}}H_{k,k'}
    +\sum_{\mathcal{K}\neq\mathcal{L}}\sum_{k\in\mathcal{K},l\in\mathcal{L}}H_{k,l},
\end{equation}
where $H_{k,l}\ll H_{k,k'}$ whenever $k,k'\in\mathcal{K}\neq\mathcal{L}\ni l$, and we have $X_1\sim \Lambda_{\mathrm{ind}}$, $X_2\sim\Lambda$, where
\begin{align}
    \Lambda_{\text{ind}}&=\max_{\mathcal{L}}\max_{l\in\mathcal{L}}\sum_{\mathcal{K}}\sum_{k\in\mathcal{K}}\|H_{k,l}\|,\\
    \Lambda&=\sum_{\mathcal{K},\mathcal{L}}\sum_{k\in\mathcal{K},l\in\mathcal{L}}\|H_{k,l}\|.
\end{align}
Alternatively, we can apply a partial Trotter decomposition without splitting the terms within each cluster~\cite{Peng20Simulating}.
This reduces the Trotter error to instead scale with $X_2\sim\Lambda_{\text{int}}$, where
\begin{equation}
    \Lambda_{\text{int}}=\sum_{\mathcal{K}\neq\mathcal{L}}\sum_{k\in\mathcal{K},l\in\mathcal{L}}\|H_{k,l}\|
    \ll\Lambda. 
\end{equation}
Each cluster can then be simulated using either product formulas or more advanced quantum algorithms.
We expect that such a hybrid approach can improve the runtime of our approach, but a detailed study of such an improvement is out of the scope of the present paper and will be left as a subject for future work.

To analyze how the error of quantum simulation affects the accuracy of the overlap estimation, we use the block-diagonal structure of Eqs.~\ref{eq:controlled_evolution1} and \ref{eq:controlled_evolution2}.
We see that the controlled time evolution has error at most $\eta$ provided that quantum simulation is performed with accuracy $\eta$.
To achieve an accuracy of $\epsilon$ in the estimate of overlap, we set $\eta_a=\mathcal{O}(\epsilon/\lambda_a)$ for $a=0,1$, respectively.
This sets the minimum number of Trotter steps $R$ in Eq.~\ref{eq:def_Trotter_number}.

We now explain how to add the double-control by the $|l\rangle$ and $|t\rangle$ registers to a general product formula $S_p$.
(The single-control by the $|t\rangle$ register also required for the SEL$_a$ oracle can be implemented by the following techniques as well.)
The near-linear dependence of the evolution time in $R$ and requirement that $L\gg \mathcal{O}(t^2)$ imply that we need $L\gg R$. This in turn implies that $lt/L$ is not necessarily an integer multiple of $t/R$.
For simplicity, we assume that $L$, $R$ are powers of 2, and write $lt/L=rt/R+q$ for $0\leq q < t/R$.
We write $q'=qR/t<1$, and then $lR/L=r+q'$ gives the number of integer ($r$) Trotter steps and the fractional remainder $q'$.
Because $L$ and $R$ are powers of two, these integers $r$ and $q'$ are already stored in the first $\log(R)$ and the last $\log(L)-\log(R)$ bits of the $l$ register, and may be identified by renaming $l$ as $(r,q')$. The integer part ($r$) determines the number of times for which $S_p(t/R)$ needs to be applied: controlling $S_p^{2^{b_r}}(t/R)$ by the $b_r$-th bit of the $|r\rangle$ control register and the $|t\rangle$ register implements the unitary
\begin{equation}
    \sum_{r,t}|r\rangle|t\rangle\langle t|\langle r|S_p^r(t/R),
\end{equation}
where here the $|t\rangle$ register dictates the angle of rotation of each component of the product formula.
For example, we would directly implement $\sum_t|t\rangle\langle t|e^{it\theta Z_i}$ bit-wise, using the $b_t$th bit of the $t$ register to control a rotation by $e^{i2^{b_t}/2^{n_t}\theta Z_i}$.
This has a gate complexity polylogarithmic in the input parameters, and so we neglect it.
We can similarly implement the final fractional Trotter step controlled by the $|q'\rangle$ register; i.e, we implement the unitary
\begin{equation}
    \sum_{q',t}|q'\rangle|t\rangle\langle t|\langle q'|S_p(tq'/R).
\end{equation}
This also has a similar cost that is polylogarithmic in the input parameters. 
The final scaling of our doubly-controlled time evolution is then identical up to logarithmic factors to the cost of implementing the Trotter evolution without control.

Our above implementation of controlled quantum simulation is developed and optimized specifically for product formulas. Another possible circuit implementation that works for not only product formulas but also more advanced quantum simulation algorithms is to use a binary representation of the evolution time and simulate for time $2^k$ with integer $k$; see \cite{childs2017quantum} for details. In any case, the complexity only scales logarithmically with the input parameters and the overhead is negligible. This justifies the comparison between the oracular models in Eq.~\ref{eq:FT_scaling_general} and Eq.~\ref{eq:NISQ_scaling_general}, as long as the PREPARE and controlled-$W_x$ circuits have lower costs than SELECT.

A naive implementation of the Trotter steps requires that we exponentiate all the terms in the Hamiltonian. For the clustered model in Eq.~\ref{eq:def_cluster}, this implies a gate complexity of $\mathcal{O}\left(N^2\right)$ to implement each Trotter step. However, this may be improved by truncating Hamiltonian terms with very small magnitudes or by switching to an advanced quantum simulation algorithm.
We also need to synthesize the rotation gates with respect to a fault-tolerant gate set, but the overhead in the circuit synthesis is asymptotically negligible.

The PREP$_0$ and PREP$_1$ oracles require loading the coefficients in Eqs.~\ref{eq:control_state1} and~\ref{eq:control_state2} respectively onto a quantum register.
As we have chosen a uniform integration measure, the integration weights for both PREP$_0$ and PREP$_1$ are independent of the value of the $l$-register, which may be prepared by simply applying a Hadamard gate to all qubits.
The remainder of our PREPARE oracles relies heavily on the QROM and the coherent alias sampling (CAS) technique of Ref.~\cite{Babbush18Encoding}, which can be used to perform the mappings $|j\rangle|0\rangle\rightarrow |j\rangle|a_j\rangle$ and $\frac{1}{\sqrt{N_d}}\sum_j|j\rangle\rightarrow\frac{1}{\sqrt{\sum_ja_j}}\sum_j\sqrt{a_j}|j\rangle$ with $\widetilde{\mathcal{O}}(N_d)$ Toffoli gates, where $N_d$ is the number of unique datapoints or indices $j$.
This is important as we do not assume that our times $t$ are chosen uniformly, so preparing the $|t\rangle$ register is non-trivial.
If we index our times by some uniform index $j$, i.e. writing $t=t_j$, we can use QROM to map $|j\rangle|0\rangle\rightarrow |j\rangle|t_j\rangle$.
(This requires the $|j\rangle$ register be of size $n_d=\log(N_d)$.)
Coherent alias sampling allows us to prepare the state $\frac{1}{\sqrt{\lambda_1}}\sum_{j,x}\frac{\sqrt{t_j}}{\sigma_{x,t_j}}|j\rangle|t_j\rangle|x\rangle$, with a cost equal to the number of unique datapoints.
Combining this with the prepared $|l\rangle$ register above yields the PREP$_0$ oracle.
If $\sigma_{x,t}=\sigma_t$ (i.e. all separate experiments are performed with the same error, which is a reasonable assumption), this has an identical cost of $N_d$.
We assume that $N_d$ scales at worst linearly in $T$ (i.e. for dense sampling), and so the cost of implementing the PREP$_0$ oracle is bounded by $\widetilde{\mathcal{O}}(T)$ and dominated in the block encoding by the additive cost of the SEL$_0$ oracle.
The PREP$_1$ oracle differs from the PREP$_0$ oracle only by the additional amplitudes $a_{x,k}$.
These may be mapped onto the device using coherent alias sampling at a cost scaling as $N_xN_{\omega}$.
This cost is additive to the $N_d$ cost above, and as we expect $N_xN_{\omega}\ll N_d$, we expect this oracle to also be dominated by the cost of SEL$_1$.

As an additional part of the SEL$_1$ subroutine, we need to implement the controlled $Z$ rotation $e^{-iZ_2(t\omega_{x,k}+\phi_x)}$. The classical values $\omega_{x,k}$ and $\phi_x$ here need to be loaded onto the quantum device.
In order to do this, we rely on the QROM technique of Ref.~\cite{Babbush18Encoding}.
Given the set of $N_{\omega}$ classical datapoints $f_{x,k}=2\pi\omega_{x,k}$ to some fixed precision, QROM can be used to perform the mapping $|k\rangle|x\rangle|0\rangle\rightarrow|k\rangle|x\rangle|f_{x,k}\rangle$, using $\widetilde{\mathcal{O}}(N_{\omega}N_x)$ Toffoli gates.
We can similarly map $\phi_x$ onto a single qubit, $|x\rangle|0\rangle\rightarrow|x\rangle|b_x\rangle$ where $b_x=0$ if $\phi_x=0$ and $b_x=1$ if $\phi_x=\pi/2$, at a cost of $\widetilde{\mathcal{O}}(N_x)$ Toffoli gates.
These mappings are more appropriate to implement during the PREP$_1$ step, so we shall insert them there.
Then, in the SEL$_1$ subroutine, we may assume access to these registers, in which case the controlled $Z$ rotation may be implemented by arithmetic of the same form as in product formulas at a cost polylogarithmic in the size of the $|f_{x,k}\rangle$ and $|t\rangle$ registers.
We may alternatively use the phase gradient method described in \cite{Sanders20Compilation}.

It remains to describe a preparation scheme for $\rho_x=W_x|0\rangle\langle 0|W_x^{\dag}$.
This is especially important to consider as $\rho_x$ is not a pure state, so it is impossible to prepare it from an initial register with a unitary operation on an $N$-qubit quantum register.
We require $W_x$ to be a unitary operation for the expectation value estimation algorithm, as it requires repeated access to $W_x$ and $W_x^{\dag}$ (or equivalently the ability to reflect around $\rho_x$).
To solve this problem, we expand the size of our system register, and prepare a purified state $|\psi_x\rangle =W_x|0\rangle$ such that for all observables $O$ within our original system, $\mathrm{Trace}[O\rho_x]=\mathrm{Trace}[O\otimes I|\psi_x\rangle\langle\psi_x|]$.
This requires that we at most double the number of qubits of our system $N$, and all operations other than $W_x$ and $W_x^{\dag}$ we can ignore the additional qubits.
As we will see below, for applications in NMR we are mostly interested in preparing states such as
\begin{equation}
    \rho_x = \frac{1}{2}(I + Z_{j_x}),
\end{equation}
which is the maximally-mixed state on all qubits except qubit $j_x$.
To achieve this with an additional $N$-qubits, we begin with $N$ copies of the Bell state $\frac{1}{\sqrt{2^N}}\left(\ket{00}+\ket{11}\right)^{\otimes N}$, which can be prepared using only Clifford gates.
We then perform a Toffoli gate with $x$ and $j_x$-th qubit as controls and $j_x+N$ as target, followed by a Hadamard gate on the $j_x$-th qubit controlled by $x$. Each controlled Hadamard can be implemented using a single Toffoli gate \cite[FIG.\ 17]{Lee21Even}.
This prepares the state
\begin{equation}
    |x\rangle|\psi_x\rangle = \frac{1}{\sqrt{2^{N-1}}}|0_{j_x}0_{j_x+N}\rangle\prod_{j\neq j_x}\Big(|0_{j}0_{j+N}\rangle + |1_j1_{j+N}\rangle\Big),
\end{equation}
which has our desired properties.
More generally, any thermal state of the classical $1D$ Ising model can be prepared as a $2N$-qubit thermofield double state with perfect fidelity using a depth $N/2$ circuit~\cite{Wu18Variational}.

\begin{figure*}
    \centering
    \includegraphics[width=\textwidth]{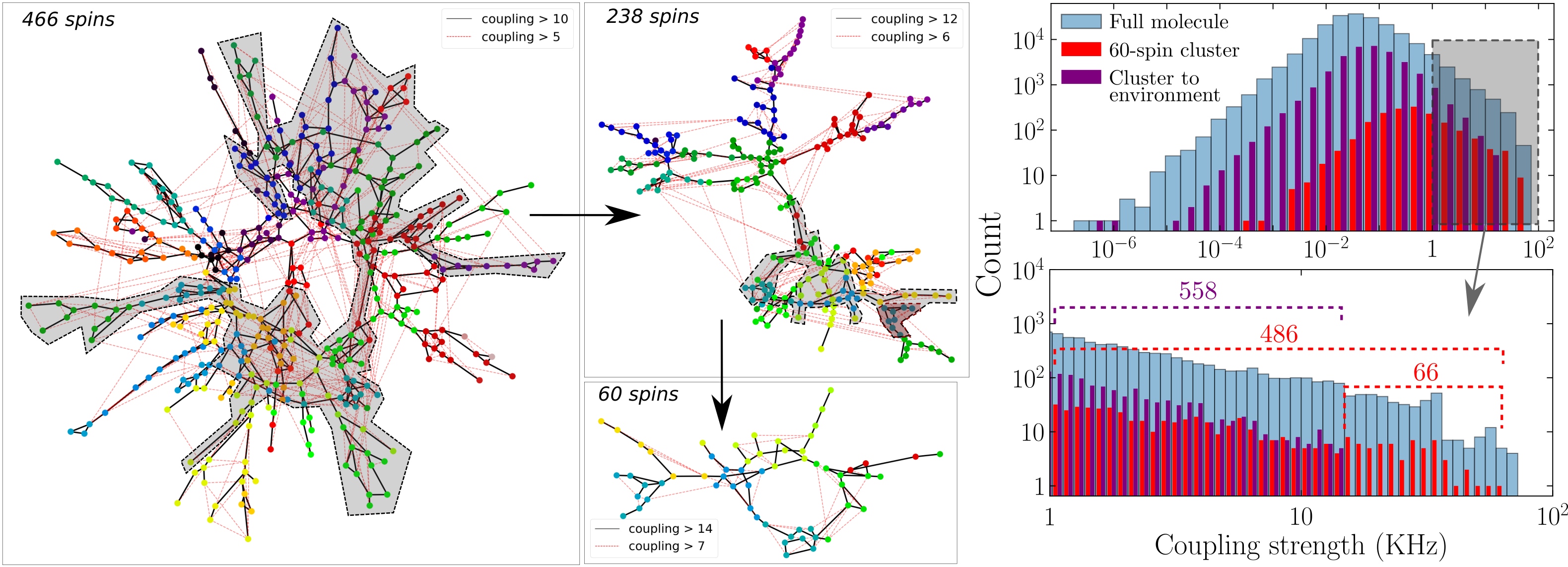}
    \caption{Finding spin clusters within the ubiquitin protein using data from Ref.~\cite{Cornilescu98Validation}. (left): a 466-spin cluster connected at a coupling of $10~\mathrm{KHz}$ (the total protein has 692 H atoms). (middle-top) a 238-spin subset of this cluster, connected at a coupling of $12~\mathrm{KHz}$. (middle-bottom) a smaller 60-spin subset of this cluster, connected at a coupling of $14~\mathrm{KHz}$. Black lines and red dashed lines between spins indicate strong and medium-strength couplings within the cluster. Red dashed region in all three clusters is an 8-spin sub-cluster studied in this text. (right-top) histogram of couplings within the 60-spin cluster and to the environment. (right-bottom) a zoom-in on the histogram tail to show the distribution of the dominant couplings.}
    \label{fig:Ubiquitin}
\end{figure*}

\section{Application to nuclear magnetic resonance spectroscopy}\label{sec:NMR}

We now focus on the application of our quantum learning algorithm to NMR spectroscopy.
In an NMR experiment, a sample of a molecule, crystal, or other material is placed in a strong magnetic field, which interacts with the magnetic moment of any spinful nucleus.
These also couple to each other through dipole-dipole and electron-mediated interactions, making the full Hamiltonian~\cite{Mcdermott12solid}
\begin{align}
    &H=\sum_i\chi_i\mathbf{S}_i\cdot\mathbf{B} + \sum_{i\neq j}J_{i,j}\mathbf{S}_i\cdot\mathbf{S}_j \nonumber\\
    &+ \sum_{i\neq j}\Gamma_{i,j}\left[\mathbf{S}_i\cdot\mathbf{S}_j-\frac{3}{|\mathbf{r}_{i,j}|^2}(\mathbf{S}_i\cdot\mathbf{r}_{i,j})(\mathbf{S}_j\cdot\mathbf{r}_{i,j})\right].\label{eq:spin_Hamiltonian}
\end{align}
Here, $\mathbf{r}_{i,j}=\mathbf{r}_i-\mathbf{r}_j$ is the vector between the $i$th and $j$th nuclei, $\mathbf{S}_i$ is the spin vector, $\mathbf{B}$ is the external magnetic field, $\chi_i=\gamma_i\delta_i$ is the shielded magnetogyric ratio of the ith nucleus (with $\delta_i$ the chemical shift and $\gamma_i$ the unshielded magnetogyric ratio), $\Gamma_{i,j}=\frac{\mu_0\gamma_i\gamma_j\hbar}{8\pi|\mathbf{r}_{i,j}|^3}$ is the dipole-dipole interaction strength (with $\mu_0$ the vacuum permeability and $\hbar$ Planck's constant), and $J_{i,j}$ is the electron-mediated coupling strength.
Assuming that our nuclei are spin-$1/2$, $\mathbf{S}_i=\frac{1}{2}(X_i,Y_i,Z_i)$.
As the coupling constants $\Gamma_{i,j}$ depend directly on the physical distance $|\mathbf{r}_{i,j}|$ between the molecular spins; knowledge of these distances is sufficient to infer the molecular geometry (modulo global translations, rotations and reflections)~\cite{Crippin78Note}. 
An NMR spectroscopy experiment follows the protocol outlined in Sec.~\ref{sec:learning_problem}.
The system begins at thermal equilibrium, is perturbed by one or more magnetic field pulses, and has its free-induction decay read out (which is equivalent to measuring $\sum_iX_i$).
In this formalism the starting state $\rho_x$ and measurement operator $O_x$ are the same for all experiments (which differ in their choice of $H_x(t)$).
However, the external perturbation is often chosen to polarize the initial state and flip spins in the end, making the signal $S_x(t)$ a local spin-spin correlation measurement.
(We will discuss how to implement this in a strongly correlated system shortly.)

Whether a NMR Hamiltonian lies within the ergodic, classically-feasible, or quantum-feasible regimes in Fig.~\ref{fig:learnability} depends on the relative energy scales of the terms in Eq.~\ref{eq:spin_Hamiltonian}.
These are typically
\begin{equation}
    \beta^{-1} >> B\chi >> B|\chi_i-\chi_j| \sim \Gamma_{i,j} >> J_{i,j},\label{eq:energy_scales}
\end{equation}
where $\beta$ is the inverse temperature of the system, and $\chi=\frac{1}{N}\sum_i\chi_i$.
Only terms coupled to the dipolar term ($\Gamma_{i,j}$) and the electron-mediated interaction ($J_{i,j}$) generate classically challenging dynamics, so one of them must be large for us to lie outside the classically feasible region of Fig.~\ref{fig:learnability}.
However, in solution a molecule tumbles rapidly, averaging out the dipolar term to $0$.
In a strong magnetic field, the remaining Hamiltonian can be treated perturbatively and solved classically, rendering our quantum learning algorithm unnecessary.
Previous proposals~\cite{Sels19Quantum} that suggested using quantum computers to learn the structure of molecules in solution suffer from this applicability issue.
(Note that in zero- or ultra-low- field NMR~\cite{Ledbetter09Optical,Theis11Parahydrogen,Blanchard13High,Barskiy19Zero} Eq.~\ref{eq:energy_scales} does not hold; in these experiments $B\chi\lesssim J_{i,j}$ and quantum computers may have a role to play in learning structure.)

We propose to instead study systems where molecules are out of solution and not free to move.
Indeed, NMR is used in a wide range of systems out-of-solution: gels~\cite{Nonappa16Solid}, surfaces~\cite{Deng08Solid}, proteins in membranes~\cite{Agarwal14Denovo,Sinnaeve20Selective} and in the solid-state~\cite{Mcdermott12solid}.
In these systems the uniform magnetic field interaction term $B\chi$ is still the dominant energy scale, and terms that do not commute with this cancel out, leaving
\begin{align}
    H=&B\sum_{i}\chi_iZ_i + \sum_{i\neq j}J_{i,j}\mathbf{S}_i\cdot\mathbf{S}_j\nonumber\\
    &+\sum_{i\neq j}\Gamma_{i,j}(3\cos^2(\phi_{i,j})-1)\left[\mathbf{S}_i\cdot\mathbf{S}_j-3Z_iZ_j\right],\label{eq:ham_secular_approx}
\end{align}
where $\phi_{i,j}$ is the angle between $\mathbf{B}$ and $\mathbf{r}_{i,j}$.
(This is commonly known as the secular approximation.)
This approximation requires $\phi_{i,j}$ to be well-defined (e.g. by stacking membranes so that all proteins are similarly aligned with the magnetic field).
Simulating randomly scattered molecules would require classical averaging over many such alignments, presenting an additional simulation challenge (and in practice broadening out spectral lines).
In solid state NMR experiments one typically removes the dipolar coupling by magic angle spinning; spinning at a high frequency around an angle $\theta=54.74^\circ$ to the magnetic field.
This spinning suppresses the dipolar term by a factor $(3\cos^2(\theta)-1)\sim 0$ as long as the spinning frequency is much higher than the dipolar coupling strength.
Combining this with frequency-selective dipolar recoupling~\cite{Gullion08Rotational} has achieved remarkable success in biochemistry, (see e.g. Ref.~\cite{Colvin16Atomic}).
However, this is not typically achievable to high precision for proton NMR, where dipolar couplings are typically of the order of $30-40$ KHz~\cite{Vinogradov99High} (by comparison, the highest frequency centrifuges are around $100$ KHz~\cite{Agarwal14Denovo}).
Even when magic-angle spinning is combined with decoupling pulse schemes, significant residual coupling in proton-NMR spectra can be observed~\cite{Lesage03Experimental,Elena04Direct,Agarwal14Denovo}.
(Moreover, suppressing the dipolar coupling term removes valuable structural information about a system.)
Proton NMR in solutions has achieved great success in biochemistry, but the folding of proteins can be quite different in vitro versus in vivo.
For example, a large number of proteins in cell walls and membranes are folded precisely according to this external environment, and lose their shape in solution~\cite{Agarwal14Denovo,Sinnaeve20Selective}.
This suggests that learning the structure of proteins in membranes via proton NMR is a potentially valuable beyond classical quantum computing application.

Proposing to study systems with strong dipolar coupling presents a challenge in designing realistic state preparation and measurement schemes.
As the temperature is larger than all energy scales in our experiment, our initial state is a thermal state
\begin{equation}
    \rho_{\mathrm{th}}=e^{-\beta H}\sim 1-\sum_i\beta B\chi_iZ_i,
\end{equation}
and we must perturb this in order to generate any signal at all.
(The approximation here is quite good, as even for a $23.5$~T background magnetic field $\beta B\chi\sim 10^{-4}$.)
We have two external handles to perturb our system: the ability to apply time-dependent RF pulses to modulate the background magnetic field $B$, and magic-angle spinning~\cite{Andrew58Nuclear, Lowe59Free}.
The latter is a double-edged sword; we cannot alter the direction of the magnetic field nor the spinning sample on the timescale of our system, so it is only practical to use this to suppress the dipolar coupling by a fixed amount for the entire experiment.
However this may be adjusted during the experiment via dipolar decoupling and recoupling pulse schemes~\cite{Levitt86Composite,Bielecki89Frequency,Paravastu06Frequency,Tycko10Homonuclear,Gullion08Rotational}.
As a simple example, consider the 4-step WAHUHA scheme~\cite{Choi20Robust,Waugh68Approach}, which consists of (1) a rest of time $dt$ and a $\frac{\pi}{2}$ pulse around the $x$ axis, (2) a rest of time $dt$ and a $-\frac{\pi}{2}$ pulse around the $y$ axis, (3) a rest of time $2dt$ and a $\frac{\pi}{2}$ pulse around the $y$ axis, and (4) a rest of time $dt$, a $\frac{-\pi}{2}$ pulse around the $x$ axis and a final rest of time $dt$.
By performing a Magnus expansion to first order in $dt$ on the above combined unitary scheme, the dipolar term averages to zero.
(This is true of any scheme that rotates the $X,Y,Z$ Pauli spins to the $z$-axis of the magnetic field for equal periods of time.)
Once the dipolar field is sufficiently suppressed, individual spins may be flipped by e.g. applying a low-amplitude magnetic field oscillating at a frequency $\omega$ which addresses spins for which $B\chi_i=\omega$.
This simple scheme is likely difficult to achieve sensitivity below $\sim 1$~KHz, but this may be improved by frequency-selective pulsing schemes~\cite{Geen91Band,Levitt86Composite} such as the DANTE (Delays Alternating with Nutations for Tailored Excitation) scheme~\cite{Morris78Selective}.
After applying such a scheme, the perturbation to the system is roughly
\begin{equation}
    \rho - \rho_{\mathrm{th}}=\beta B\sum_{|\chi_iB-\omega|<\delta\omega}\chi_i Z_i,\label{eq:subtracted_state}
\end{equation}
where $\delta\omega$ gives the accuracy of the technique.
In an NMR experiment the signal is rescaled by the factor $\beta$ in Eq.~\ref{eq:subtracted_state}.
However, in a quantum computation we may divide by $\beta$ before performing our estimation, which makes the error requirements in our final signal independent of $\beta$.
Approximations in the above can be accounted for in our learning scheme by adjusting the starting state, or by incorporating the pulse sequence into the unitary $U_x$, giving an additional advantage over classical Hamiltonian inference.

\subsection{Determining the structure of ubiquitin as an example application}~\label{sec:ubiquitin}
\begin{figure}
    \centering
    \includegraphics[width=\columnwidth]{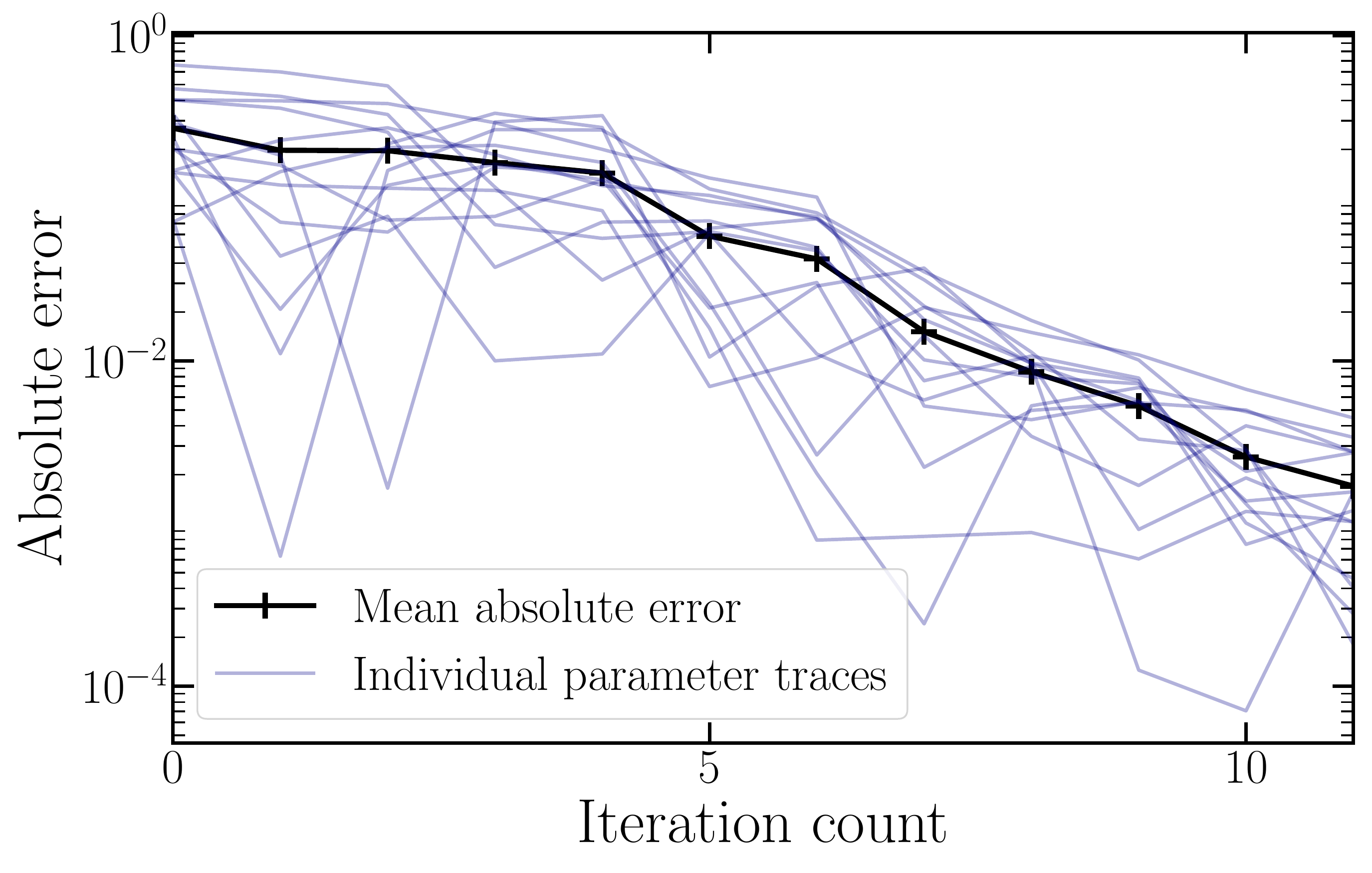}
    \caption{Learning small couplings of a 6-spin cluster within ubiquitin given a fixed backbone from a noisy dataset with standard deviation $10^{-3}$. The system is learned from a dataset of $21$ points equally spaced between $0$ and $2$~ms (assuming a $23.5$ T background magnetic field, and a dipolar suppression of $\alpha=10$). Absolute error in individual parameters (blue faint lines) and the mean absolute error (black line) is plotted at each iteration of the CG algorithm.}
    \label{fig:convergence}
\end{figure}

A significant body of literature on protein structures already exists, which we can use as a benchmark for our quantum learning algorithm.
The protein ubiquitin, named for its abundance throughout eukaryotic organisms, was discovered in 1975~\cite{Goldstein75Isolation,Wilkinson12Chain}.
The structure of ubiquitin is well-known, making it a good benchmark.
The protein contains over $600$ protons, but these tend to cluster (in terms of their relative coupling strengths, which correspond to their location in space); we propose to divide the full molecule into smaller clusters that may be studied individually.
This corresponds to assuming our experimental signature 
\begin{equation}
    S_x(t)=\sum_{\mathrm{cluster}\;c}S_{x,c}(t),\label{eq:linear_combination_signal}
\end{equation}
and we can either subtract the contribution of individual pieces from the total signal, or learn this linear combination in a single step.
The latter has a linear overhead in the number of pieces, and does not require larger quantum computers or longer circuits.
In practice we do not need to determine the clustering ahead of time; clusters will present themselves as minibands in the spectrum that cannot be separated as in Eq.~\ref{eq:linear_combination_signal}, and we may account for these in our learning algorithm by allowing our optimization to adjust the number of spins in any given band.

To investigate this clustering, we write the ubiquitin molecule as a graph (with edges weighted by the dipolar interaction strength).
We take the atomic co-ordinates of ubiquitin in solution from previous NMR data (protein data bank ID 1D3Z), and proton chemical shifts from the Biological Magnetic Resonance Data Bank (BMRB), entry 17769 (both in turn taken from Ref.~\cite{Cornilescu98Validation}).
We then define a cluster as any connected subgraph where all edges are higher weight than any edges (in the larger graph) that point from the subgraph out.
(The above definition in principle removes some edges from the subgraph / couplings from the cluster, but upon identification of the cluster we consider all couplings between the spins regardless of their size.)
Such subgraphs may be found by thresholding; setting a truncation threshold $V_{\min}$ and eliminating all couplings lower than this threshold in the nuclear spin Hamiltonian cuts the graph of spins into a set of disconnected subgraphs, each of which is a cluster.
In the ubiquitin molecule (Fig.~\ref{fig:Ubiquitin}), by adjusting this $V_{\min}$ we can first split off a 466 spin central core of the molecule (left), and then a 238-spin backbone (middle top), and finally a 60-spin cluster (middle bottom).
Given its size, we expect the problem of learning the 60-spin cluster Hamiltonian to lie around the beyond-classical boundary.
To investigate the cluster's coupling to its environment, we plot the distribution of coupling strengths (Fig.~\ref{fig:Ubiquitin}, right) both within the full molecule, within the $60$-spin cluster, and between the cluster and the rest of the molecule.
We see that the mean coupling within the ubiquitin cluster is around two to three times all couplings to the environment, and the majority of couplings to the environment are more than ten times smaller than the couplings within the cluster.
We believe that this is sufficiently weak that these couplings may be treated perturbatively, though verifying this is a clear task for future study.

We now demonstrate our Hamiltonian learning algorithm for a small cluster of 6 spins in the ubiquitin molecule (Fig.~\ref{fig:convergence}).
We assume that we have access to magic angle spinning or dipolar decoupling techniques to suppress our dipolar field by a factor $\alpha=10$, and that the background field is $23.5$~T.
To simulate the proposal that we know the protein backbone and are focused on learning long-range couplings, we start from a Hamiltonian (Eq.~\ref{eq:ham_secular_approx}) where all couplings larger than some $V_{\min}$ are known precisely, and set ourselves the task of learning smaller couplings.
This leaves 12 couplings to learn (we treat $XX+YY$ and $ZZ$ couplings independently), which we initialise at $0$.
To simulate sampling noise from the quantum computer, to each query of the device for $S_x(t)$ or the gradient we add a normally-distributed error term with a value of $10^{-3}$.
Using the conjugate gradient optimization algorithm implemented in scipy~\cite{scipy}, we find that our learning problem converges to a total error of $0.008$ KHz in only $11$ iterations (a relative error of $0.2\%$).

\begin{figure}
    \centering
    \includegraphics[width=\columnwidth]{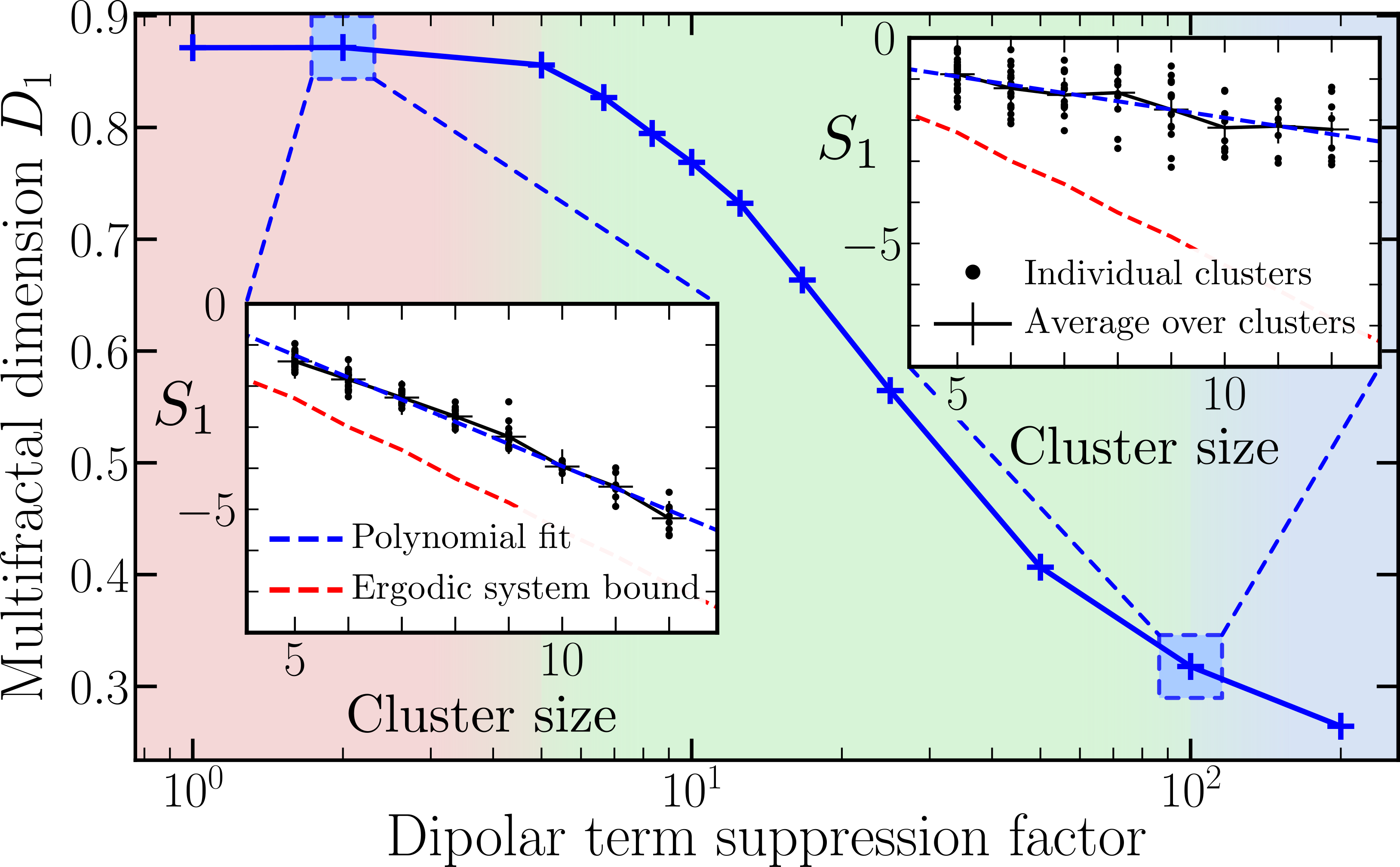}
    \caption{Plot of the multifractal dimension of small clusters of spins in the ubiquitin molecule as the dipolar term is suppressed in a background $23.5$~T field.
    Points in the main plot are extracted from a linear fit of the mean participation entropy of the different clusters as a function of the system size, as demonstrated in the two insets for a suppression factor $2$ and $100$.
    From these fits the multifractal dimension can be estimated.
    The colouring of the plot indicates the expected quantum and classical learnability of the system as the dipolar term is suppressed, corresponding to the different regions in Fig.~\ref{fig:learnability}.}
    \label{fig:multifractal_dimension}
\end{figure}

\subsection{Learnability of spin clusters in ubiquitin}

It remains to demonstrate that spin clusters in ubiquitin in a membrane or cell wall will generate an NMR signal within the region of quantum feasibility in Fig.~\ref{fig:learnability}.
In order to study this, we investigate the participation entropy~\cite{Mace19Multifractal}
\begin{equation}
    S_1=\sum_{\alpha}|\psi_{\alpha}^2|\log(|\psi_{\alpha}^2|)
\end{equation}
of eigenstates $|\psi\rangle=\sum_{\alpha}\psi_{\alpha}|\alpha\rangle$ of small clusters in the ubiquitin protein.
By adjusting our threshold $V_{\min}$ in the clustering protocol described above, we identify a collection of 127 clusters of $N=5-12$ spins in ubiquitin.
We do not expect the spectra of the ubiquitin Hamiltonian to be represented by a disjoint sum of some of these clusters (which would make it classically simulatable): the truncation is simply performed to give an ensemble on which to study, and we expect that the spectra from some of these clusters in the larger spin environment would differ significantly from those of the truncated piece.
The dipolar couplings in these Hamiltonians may be suppressed relative to their chemical shifts by magic angle spinning or decoupling pulse schemes; we simulate this numerically by dividing the dipolar term by a variable suppression factor $\alpha$.
For the Hamiltonian of each cluster (Eq.~\ref{eq:ham_secular_approx}) we calculate the mean participation entropy $S_1$ across the middle half of the spectrum at $P=N/2$ half-filling as we increase the dipolar term suppression by a factor $\alpha$.
(For odd-sized clusters of $N=5,7,9,11$ spins, we take $P=(N-1)/2$-filling.)
In an ergodic system where nearly all computational basis states contribute nearly equally to each eigenstate, the participation entropy scales as $S_1=-\log[\mathrm{dim}(\mathcal{H})]=-\log[{N\choose P}]$.
For the systems considered this is roughly $S_1^{(\mathrm{ergodic})}\sim -0.63N$.
By comparison, a completely localized system has a constant (or logarithmically-growing) participation entropy.
As we increase $\alpha$ and the system becomes non-ergodic, the participation entropy follows a trend $S_1\sim D_1S_1^{(\mathrm{ergodic})}$, where the multifractal dimension $D_1=D_1(\alpha)$ characterizes the fraction of the Hilbert space explored by an eigenstate~\cite{Mace19Multifractal}.

In Fig.~\ref{fig:multifractal_dimension} we plot the multifractal dimension of our 127 spin clusters as we suppress the dipolar term by a factor $\alpha$.
Our quantum-feasible region corresponds to a multifractal dimension $D_1<1$, while our classically-feasible region corresponds to a multifractal dimension $D_1<<1$.
We see a clear region between a suppression factor of $\alpha\sim 5$ and $\alpha\sim 100$ where our system begins to localize and the multifractal dimension quickly drops, but the system is not completely localized and classically simple.
In the thermodynamic limit this transition is discontinuous~\cite{Mace19Multifractal}, but as we are interested in finite system sizes, we believe that the observed trend of $D_1$ as we cross this localization transition is relevant to our situation.
On either side of the localization transition, local disorder will make individual clusters either more ergodic or more local than the mean, which implies that the boundaries between degenerate, quantum-feasible and classically-feasible are not sharp as a function of the dipolar suppression.
(This can be seen in the insets of Fig.~\ref{fig:multifractal_dimension}.)

\begin{figure}
    \centering
    \includegraphics[width=\columnwidth]{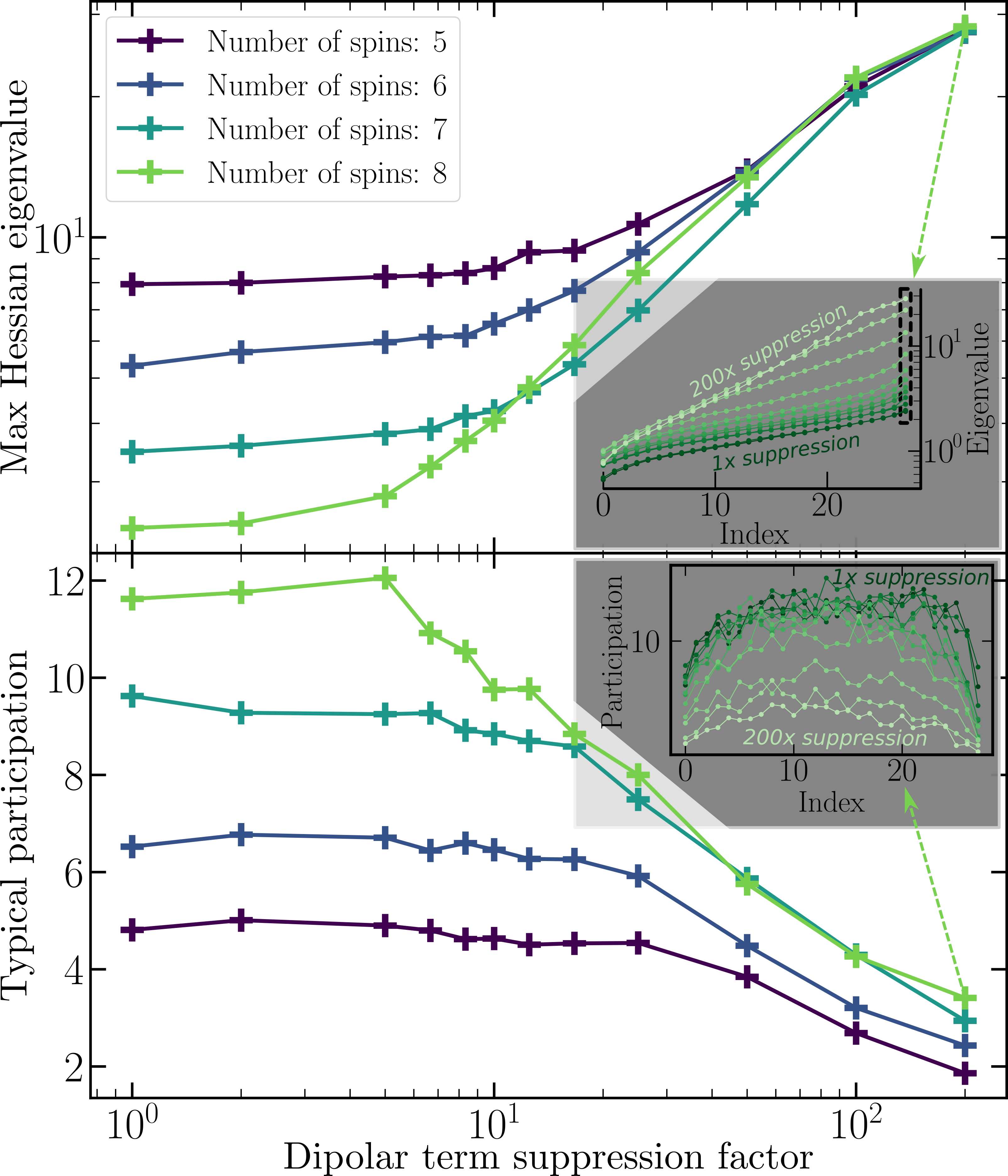}
    \caption{Eigenvalue and eigenvector participation data for Hessians of the Hamiltonian learning problem for small spin clusters in ubiquitin.
    (Data is taken as the set of $\langle Z_i(t)Z_j(0)\rangle$ and $\langle X_i(t)X_j(0)\rangle$ correlators using a set of equally spaced times between $t=0$ and $t=5$~ms in the absence of sampling noise, implying that units of the Hessian eigenvalues are arbitrary.)
    (Top) typical (geometric mean) largest Hessian eigenvalue for different system sizes as the dipolar term is suppressed. Inset shows the typical Hessian spectrum for each suppression factor over clusters of $8$ spins - the light green line in the main plot is taken from the indicated cut through the inset.
    (Bottom) typical (geometric mean) eigenvector participation (Eq.~\ref{eq:Hessian_participation}) for the same dataset. Inset shows the participation across the entire Hessian spectrum for clusters of $8$ spins; each datapoint in the light green line in the main plot corresponds to a geometric mean over a single line in the inset.}
    \label{fig:hessian_data}
\end{figure}

The ergodic to non-ergodic phase transition observable in the Hamiltonian eigenstructure maps immediately to the learnability of the NMR spin Hamiltonian.
This can be studied in the Hessian of the Hamiltonian learning problem at complete convergence, given by Eq.~\ref{eq:cost_function_second_deriv_approx}.
As the Hessian corresponds to the Fisher information of the system, small eigenvalue-eigenvector pairs $(\lambda,\vec{v})$ correspond to `floppy modes' in our parameter space; linear combinations of parameters that may be adjusted in tandem without significantly altering our signal.
Large eigenvalue-eigenvector pairs correspond to combinations of parameters that are well-learned.
In Fig.~\ref{fig:hessian_data} we study the typical eigenstructure of the Hessians of clusters of $5-8$ spins.
We see a clear transition that corresponds exactly to the ergodic to non-ergodic phase transition identified in Fig.~\ref{fig:multifractal_dimension}.
On the left-hand side, corresponding to the ergodic phase, the typical maximum eigenvalue (Fig.~\ref{fig:hessian_data}, top) of the ensemble shows a clear exponential decay in system size, implying that learning in a large system will be nigh-impossible.
Moreover, the typical participation in these systems,
\begin{equation}
    \exp\left(-\sum_jv_j^2\log{v_j^2}\right),\label{eq:Hessian_participation}
\end{equation}
grows quickly (Fig.~\ref{fig:hessian_data}, bottom), implying that these modes correspond to global data rather than specific couplings.
By contrast, when the system is strongly localized, the largest eigenvalues are roughly constant in the system size, and correspond to linear combinations of only one or two parameters.
As the system shifts between these two phases, we see a continuous improvement in learnability, where it appears we can learn some of but not all of the system.
This can be observed in the full eigenspectrum data (Fig.~\ref{fig:hessian_data}, top inset).
We note that the largest eigenvalues also correspond to smaller typical participation, which suggests that when a system is on the ergodicity boundary we can learn some local couplings rather than just global information.
This result demonstrates the importance of having access to the Hessian when solving the learning problem, as it tells which of the converged parameters can be relied upon.

\section{Conclusion}

In this work we introduced a new method for learning an unknown quantum Hamiltonian of a spin system from time-resolved measurements of the system.
We constructed and costed circuits within NISQ and FT frameworks to estimate gradients of the cost function of this learning problem, finding clear asymptotic speedups when one is not constrained by poor coherence in NISQ devices.
We outlined an application for these algorithms in classically intractable NMR experiments, and proposed a specific region in the NMR field (when dipolar couplings are strong and cannot be simply removed) as an area where beyond-classical computations may be very useful.
Taking the ubiquitin protein as an example, we identified small clusters of spins in the larger $^1$H dipolar coupling matrix, demonstrated the convergence of our learning algorithm on a toy example, and investigated the cluster-environment coupling and the learnability of the system as the dipolar coupling is suppressed.

As part of this work, we identified a direct correspondence between the ergodic to non-ergodic phase transition and the learnability of a Hamiltonian from time-resolved experimental data.
The latter was clearly observable in the structure of the Hessian of the learning problem.
We believe this to be a more general result than in the NMR problems that we have studied in this work.
As far as we know, this is the first link demonstrated between the onset of wavefunction fractality and the onset of learnability of the generating Hamiltonian.

Our work opens a new field of quantum computing applications, leaving clear directions for future study.
Our learnability data suggests that not all dipolar coupling parameters (that encode the 3D protein structure) may be learnable using the simple spin-spin correlations measured in this work.
Developing future experiments to target the parameters relevant for structure calculations in NMR systems will be highly relevant in the future.
A related question (which we have not yet determined the answer to) pertains to the conditions (if any) under which this problem is classically difficult.
While we can rely on the fact that inference of NMR spectra from strongly-correlated problems appears difficult (and the forward problem of generating the spectra is at least DQC1-hard~\cite{Somma08Parameter,Knill98Power}), we do not know of a complexity-theory result explaining when the inverse problem is as difficult.
Also, the protocol we propose in Sec.~\ref{sec:robust_learning} can, in principle, avoid trapping in local minima; however this is not optimized or fully costed.
We suggest that following Ref.~\cite{Somma08Parameter}, it may be possible to define a protocol using our methods that learns at the Heisenberg limit (when learning is possible).
It is also unclear precisely how well our protocol behaves in a system with a large number of unknown couplings, and whether we will have problems with vanishing gradients as our system size grows.
(We note however that this is avoided somewhat in NMR problems where we start with a reasonable guess of our Hamiltonian parameters.)
Another clear direction for future work is to consider learning Hamiltonian parameters from the Fourier transform of the spectra $S_x(t)$ instead of the time-resolved data, as spectral information is typically more robust to noise than amplitude data.
It also remains to determine what the cost is to optimize a real-world NMR experiment from a realistic initial guess.
Finally, thanks to the generality of our Hamiltonian learning techniques, it should be possible to extend these methods to the design of new NMR experiments, or to interpret data from other types of experimental procedures used to probe condensed matter, high energy physics, chemistry and materials science.
We look forward to further explorations of what new possibilities for quantum experiments and data analysis these techniques can bring to the scientific community.

\subsection*{Acknowledgements}

The authors wish to thank Igor Aleiner, Dmitry Abanin, Nicholas Rubin, Jarrod McClean, Joonho Lee, Ashok Ajoy, Joshua Goings, Nathan Wiebe and William Huggins for useful advice and discussions of this work. D.F. was supported by NIH grant GM065334.

\bibliography{NMR,ryan}

\appendix

\onecolumngrid

\section{Optimizing query counts for NISQ algorithms}\label{app:msmt_optimization}

In this section we derive Eq.~\ref{eq:NISQ_scaling_general} by optimizing the allocation of oracle calls across a set of NISQ experiments.
This is a slightly atypical approach; typically in a NISQ experiment one attempts to optimize the number of experiments (or shots) to achieve some target error $\epsilon$~\cite{Wecker15Towards,Rubin18Application} rather than working with an oracular model.
However, this optimization allows us to make a fair comparison between the results of Sec.~\ref{sec:algorithms} and Sec.~\ref{sec:ft}.
Propagating variance through Eq.~\ref{eq:cost_function_derivative} yields
\begin{equation}
    \mathrm{Var}\left[\frac{dC[H]}{dh_n}\right]=\sum_{x,t}\frac{1}{\sigma_{x,t}^4}\left[\mathrm{Var}\left[\bar{S}_x(t)\right]\bar{J}_x^n(t)^2 + \mathrm{Var}\left[\bar{J}_x^n(t)\right]\left[\bar{S}_x(t)-S_x(t)\right]^2\right].
\end{equation}
As our estimate of $\bar{J}_x^n(t)$ is bounded, assuming $\|O_x\|=1$ the bound in Eq.~\ref{eq:NISQ_Hoeffding_bound} yields an estimator for $\bar{J}_x^n(t)$ with variance $\epsilon^2$ using $M=t^2\epsilon^{-2}$ repetitions of the circuit.
We now assume that we may measure $S_x(t)$ and $J_x^n(t)$ for different $x$ in parallel.
This is realistic for our NMR application, and will save a factor $N_x$ in the asymptotic scaling.
As the number of oracle calls per circuit scales as $t$, we can achieve a variance $\mathrm{Var}\left[\bar{J}_x^n(t)\right]=a_{x,t,J}t^3C_{t,J}^{-1}$ using $C_{t,J}$ oracle calls (for some $t$ and $x$-independent constant $a_{x,t,J}$).
Estimating $\bar{S}_x(t)$ with to a variance $\epsilon^2$ requires simply repeating the corresponding circuit (Fig.~\ref{fig:NISQ_circuits}, top) $\epsilon^{-2}$ times, and so with $C_{t,S}$ oracle calls we achieve a variance $\mathrm{Var}\left[\bar{S}_x(t)\right]=a_{x,t,S}tC_{t,S}^{-1}$ (for some $t$ and $x$-independent constant $a_{x,t,S}$).
As $[\bar{S}_x(t)-S_x(t)]$ is independent of $t$ (being bounded by $2$ when $\|O_x\|=1$), and $\bar{J}_x^n(t)$ scales linearly in $t$ (as discussed in the main text), assuming that $\sigma_{x,t}=\sigma$ we have
\begin{equation}
    \mathrm{Var}\left[\frac{dC[H]}{dh_n}\right]=\sum_{x,t}\frac{1}{\sigma^4}\left[\frac{a_{x,t,S}t^3}{C_{t,S}}+\frac{a_{x,t,J}t^3}{C_{t,s}}\right],
\end{equation}
where we have absorbed the constants of the $\bar{S}_x(t)$ and $\bar{J}_x^n(t)$ scaling into $a_{x,t,S}$ and $a_{x,t,J}$.
To optimize this, we adopt the same Lagrangian approach introduced for measurement optimization in Ref.~\cite{Rubin18Application}.
We write a Lagrangian
\begin{equation}
    \mathcal{L} = \lambda\left\{\sum_{x,t}\frac{1}{\sigma^4}\left[\frac{a_{x,t,S}t^3}{C_{t,S}}+\frac{a_{x,t,J}t^3}{C_{t,J}}\right]-\mathrm{Var}\left[\frac{dC[H]}{dh_n}\right]\right\} + \sum_{x,t}(C_{x,t,J}+C_{x,t,S}),
\end{equation}
and then differentiate with respect to our free parameters $C_{x,t,J}$ and $C_{x,t,S}$ and solve for the result being equal to $0$.
\begin{align}
    \frac{\partial\mathcal{L}}{\partial C_{t,S}}&= 1-\frac{\sum_x\lambda a_{x,t,S}t^3}{\sigma^4C_{t,S}^2}=0\rightarrow C_{t,S}=\lambda^{\tfrac{1}{2}}\left[sum_xa_{x,t,S}\right]^{\tfrac{1}{2}}t^{\tfrac{3}{2}}\sigma^{-2}\\
    \frac{\partial\mathcal{L}}{\partial C_{t,J}}&= 1-\frac{\sum_x\lambda a_{x,t,J}t^3}{\sigma^4C_{t,J}^2}=0\rightarrow C_{t,J}=\lambda^{\tfrac{1}{2}}\left[\sum_x a_{x,t,J}\right]^{\tfrac{1}{2}}t^{\tfrac{3}{2}}\sigma^{-2}.
\end{align}
Substituting into the expression for a variance $\mathrm{Var}\left[\frac{dC[H]}{dh_n}\right]=\epsilon^2$ yields
\begin{equation}
    \epsilon^2=\lambda^{-\tfrac{1}{2}}\sigma^{-2}\sum_{t}\left\{\left[\sum_xa_{x,t,J}\right]^{\tfrac{1}{2}}t^{\tfrac{3}{2}}+\left[\sum_xa_{x,t,S}\right]^{\tfrac{1}{2}}t^{\tfrac{3}{2}}\right\},
\end{equation}
and rearranging for $\lambda obtains$
\begin{equation}
    \lambda=\epsilon^{-4}\sigma^{-4}\left\{\sum_{t}\left(\left[\sum_x a_{x,t,J}\right]^{\tfrac{1}{2}}t^{\tfrac{3}{2}}+\left[\sum_xa_{x,t,S}\right]^{\tfrac{1}{2}}t^{\tfrac{3}{2}}\right)\right\}^2.
\end{equation}
Finally, we can write the total number of oracle calls as
\begin{equation}
    C=\sum_{x,t}(C_{x,t,S}+C_{x,t,J}) = \epsilon^{-2}\sigma^{-4}\left\{\sum_{t}\left(\left[\sum_xa_{x,t,J}\right]^{\tfrac{1}{2}}t^{\tfrac{3}{2}}+\left[\sum_xa_{x,t,S}\right]^{\tfrac{1}{2}}t^{\tfrac{3}{2}}\right)\right\}^2,\label{eq:final_cost}
\end{equation}
as $a_{x,t,J}$ and $a_{x,t,S}$ are constants, we have asymptotically that $\sum_xa_{x,t,J},\sum_xa_{x,t,S}\sim \mathcal{O}(N_x)$.
Substituting this into Eq.~\ref{eq:final_cost} yields immediately Eq.~\ref{eq:NISQ_scaling_general}.

\section{Fault-tolerant approach to the backwards problem}\label{app:FT_detailed}

In this appendix, we expand on the circuitry and cost analysis of the fault-tolerant estimation of the gradient of the Hamiltonian learning problem in Sec.~\ref{sec:ft}.

We consider estimating the following quantity
\begin{equation}
\label{eq:backward}
    \sum_{x=1}^{N_x}\sum_{j=1}^{N_d}\frac{1}{\sigma_{x,t_j}^2}
    \left(\mathrm{Trace}\left(O_x \widetilde{\rho}_x(t_j)\right)
    -\sum_{m=1}^{N_\omega}a_{x,m}\cos(t_j\omega_{x,m}+\phi_x)\right)
    \int_{0}^{t_j}ds\ \mathrm{Trace}\left(O_x\left[V_n(t_j,s),\widetilde{\rho}_x(t_j)\right]\right)
\end{equation}
on a fault-tolerant quantum computer for the backwards problem. Here, operators $V$ and $\widetilde{\rho}_x$ are defined as
\begin{equation}
    V_n(t,s):=e^{-i(t-s)H}V_ne^{i(t-s)H},\qquad
    \widetilde{\rho}_x(t):=e^{-itH}\ketbra{\psi_x}{\psi_x}e^{itH}
\end{equation}
for some Hamiltonian $H$, an arbitrary term $V_n$ in $H$ and $x$-dependent initial state $\ket{\psi_x}$.

Our approach depends on a quantum overlap estimation algorithm which we review in \append{qoe}. To implement this on a fault-tolerant quantum computer, we need to discretize the time integral and re-express the target quantity as a linear combination of quantum overlaps. We also need to truncate real parameters $\sigma_{x,t_j}$, $t_j$, $a_{x,m}$, and $\omega_{x,m}$ to a finite number of digits to construct the quantum circuit. We analyze the truncation and discretization error in \append{error}. 

\subsection{Quantum overlap estimation}
\label{append:qoe}
To estimate \eq{backward} on a fault-tolerant quantum computer, we first re-express it as a linear combination of quantum overlaps:
\begin{align}
    &\sum_{x,j}\frac{1}{\sigma_{x,t_j}^2}\bra{\psi_x}e^{it_jH}O_x e^{-it_jH}\ket{\psi_x}
    \int_{0}^{t_j}ds\left(\bra{\psi_x}e^{it_jH}O_xe^{-i(t_j-s)H}V_ne^{-isH}\ket{\psi_x}
    -\bra{\psi_x}e^{isH}V_ne^{i(t_j-s)H}O_xe^{-it_jH}\ket{\psi_x}\right)\nonumber\\
    &+\sum_{x,j}\frac{1}{\sigma_{x,t_j}^2}\sum_{m}a_{x,m}\cos(t_j\omega_{x,m}+\phi_x)
    \int_{0}^{t_j}ds\left(\bra{\psi_x}e^{it_jH}O_xe^{-i(t_j-s)H}V_ne^{-isH}\ket{\psi_x}
    -\bra{\psi_x}e^{isH}V_ne^{i(t_j-s)H}O_xe^{-it_jH}\ket{\psi_x}\right).
\end{align}
Then our goal is to find unitary operators $\sel_a$, quantum states $\ket{\Psi^{a}}=\prep_a\ket{0}$, and positive numbers $\lambda_a>0$ for $a=0,1$, such that $\bra{\Psi^a}\sel_a\ket{\Psi^a}$ compute the desired linear combinations up to scaled-down factors of $\lambda_a$; equivalently, we say the target quantities are block-encoded by $\ket{\Psi^a}$ and $\sel_a$ with scaled-down factor $\lambda_a$. We describe how to construct such $\sel_a$ and $\ket{\Psi^a}$ in Sec.~\ref{sec:ft}.

We estimate the quantum overlap $\bra{\Psi}\sel\ket{\Psi}$ using the overlap estimation algorithm of Ref.~\cite{Knill07Optimal}. Specifically, we consider the two reflections
\begin{equation}
    I-2\ketbra{\Psi}{\Psi},\qquad
    I-2\sel\ketbra{\Psi}{\Psi}\sel^\dagger.
\end{equation}
These reflections keep the two-dimensional subspace $\mathrm{span}\{\ket{\Psi},\sel\ket{\Psi}\}$ invariant, on which their product
\begin{equation}
    \left(I-2\ketbra{\Psi}{\Psi}\right)
    \left(I-2\sel\ketbra{\Psi}{\Psi}\sel^\dagger\right)
\end{equation}
has eigenvalues $e^{\pm i2\arccos\abs{\bra{\Psi}\sel\ket{\Psi}}}$. Therefore, we can perform quantum phase estimation on the above operator and take the cosine of the outcome to estimate the amplitude $\alpha\sim\abs{\bra{\Psi}\sel\ket{\Psi}}$. To further retrieve the phase, we introduce an ancilla qubit and estimate
\begin{align}
    \beta_0&\sim\abs{\bra{+\Psi}\mathrm{c-}(\sel)\ket{+\Psi}}=\frac{\abs{1+\bra{\Psi}\sel\ket{\Psi}}}{2},\nonumber\\
    \beta_{\pi/2}&\sim\abs{\bra{+\Psi}\left(e^{i\frac{\pi}{4}Z}\otimes I\right)\mathrm{c-}(\sel)\ket{+\Psi}}=\frac{\abs{1-i\bra{\Psi}\sel\ket{\Psi}}}{2},
\end{align}
where $\mathrm{c-}(\sel)$ is the controlled operation
\begin{equation}
    \mathrm{c-}(\sel):=\ketbra{0}{0}\otimes I+\ketbra{1}{1}\otimes \sel.
\end{equation}
From these we obtain
\begin{equation}
    y=\frac{4\beta_0^2-\alpha^2-1}{2}+i\frac{4\beta_{\pi/2}^2-\alpha^2-1}{2}\sim\bra{\Psi}\sel\ket{\Psi}.
\end{equation}

The standard quantum phase estimation outputs an estimate of the eigenphase with accuracy $\epsilon$ by making $\mathcal{O}(1/\epsilon)$ queries to the reflections, succeeding with a constant probability greater than $1/2$. The precision parameter $\epsilon$ directly translates to a maximal error of $O(\epsilon)$ in the estimated overlap. To succeed with a higher probability, we can repeat quantum phase estimation and take the median of the outcomes. By Hoeffding's inequality, the success probability can be made arbitrarily close to one with only logarithmic overhead. We thus obtain:
\begin{lemma}[Quantum overlap estimation]
\label{lem:qoe}
    Given quantum state $\ket{\Psi}=\prep\ket{0}$, unitary $\sel$, $\epsilon>0$, and $0<\delta<1$, there exists a quantum algorithm with output $y$ such that
    \begin{equation}
        \mathbb{P}\left(\abs{y-\bra{\Psi}\sel\ket{\Psi}}\geq \epsilon\right)<\delta.
    \end{equation}
    This algorithm makes $\mathcal{O}(\log(1/\delta)/\epsilon)$ queries to $\prep$ and $\sel$ (or their controlled version $\mathrm{c-}(\prep)$ and $\mathrm{c-}(\sel)$), and uses $\mathcal{O}(N\cdot\mathrm{polylog}(1/\epsilon,1/\delta))$ additional gates, where $N$ is the number of qubits in the target system.
\end{lemma}

In the description of the above algorithm, we have ignored the normalization factor $\lambda>0$ introduced by $\ket{\Psi}=\prep\ket{0}$ and $\sel$. To get the target quantity, we need to multiply the outcome of quantum overlap estimation by $\lambda$. To ensure that the estimation succeeds with probability $1-\delta$ and accuracy $\epsilon$, it then suffices to make $\mathcal{O}(\log(1/\delta)\lambda/\epsilon)$ queries to $\prep$ and $\sel$ and use $\mathcal{O}(N\cdot\mathrm{polylog}(\lambda,1/\epsilon,1/\delta))$ additional gates.

It is instructive to compare the fault-tolerant approach with the sampling-based approach that is more suitable to implement on near-term quantum devices. That approach uses the generalized Hadamard test which produces an unbiased estimate of the real and imaginary part of $\bra{\Psi}\sel\ket{\Psi}$ with constant variance. By Hoeffding's inequality, it suffices to take $\mathcal{O}(\log(1/\delta)\lambda^2/\epsilon^2)$ samples to estimate with accuracy $\epsilon$ and probability $1-\delta$. Each sample requires constant number of queries to $\prep$ and $\sel$. Therefore, we get a factor of $\Theta(\lambda/\epsilon)$ saving by switching to the fault-tolerant quantum algorithm. See Sec.~\ref{sec:algorithms} and \ref{sec:ft} for further discussions of these two approaches.

\subsection{Truncation and discretization error}
\label{append:error}
In this section, we analyze the error due to the truncation of real parameters and discretization of the time integral. We will see in Sec.~\ref{sec:ft} that this only introduces a logarithmic overhead in the overall cost.

We first consider discretizing the integral as
\begin{equation}
    \int_{0}^{t_j}ds\ f(s)\approx\frac{t_j}{L}\sum_{\ell=0}^{L-1}f\left(\frac{\ell}{L}t_j\right),
\end{equation}
where
\begin{equation}
    f(s):=\mathrm{Trace}\left(O_x\left[e^{-i(t_j-s)H}V_ne^{i(t_j-s)H},e^{-it_jH}\ketbra{\psi_x}{\psi_x}e^{it_jH}\right]\right).
\end{equation}
This discretization error can be made arbitrarily small by choosing $L$ sufficiently large.
Here, we analyze how the error scales as a function of $L$. Using the integral expansion
\begin{align}
    \int_{0}^{t_j}ds\ f(s)-\frac{t_j}{L}\sum_{\ell=0}^{L-1}f\left(\frac{\ell}{L}t_j\right)
    =&\sum_{\ell=0}^{L-1}\int_{\frac{\ell}{L}t_j}^{\frac{\ell+1}{L}t_j}ds\left(f(s)-f\left(\frac{\ell}{L}t_j\right)\right)\nonumber\\
    =&\sum_{\ell=0}^{L-1}\int_{\frac{\ell}{L}t_j}^{\frac{\ell+1}{L}t_j}ds\int_{\frac{\ell}{L}t_j}^{s}d\tau\ f'(\tau),
\end{align}
we have
\begin{equation}
    \left|\int_{0}^{t_j}ds\ f(s)-\frac{t_j}{L}\sum_{\ell=0}^{L-1}f\left(\frac{\ell}{L}t_j\right)\right|
    \leq\sum_{\ell=0}^{L-1}\int_{\frac{\ell}{L}t_j}^{\frac{\ell+1}{L}t_j}ds\int_{\frac{\ell}{L}t_j}^{s}d\tau\norm{f'}_{\max}
    =\frac{t_j^2}{2L}\norm{f'}_{\max},
\end{equation}
where
\begin{equation}
    \norm{f'}_{\max}:=\max_{0\leq s\leq t_j}\abs{f'(s)}
\end{equation}
is the largest derivative of $f$ within the time interval $[0,t_j]$. The derivative $f'(s)$ takes the form
\begin{equation}
    f'(s):=\mathrm{Trace}\left(O_x\left[e^{-i(t_j-s)H}\left[iH,V_n\right]e^{i(t_j-s)H},e^{-it_jH}\ketbra{\psi_x}{\psi_x}e^{it_jH}\right]\right),
\end{equation}
which gives
\begin{equation}
    \norm{f'}_{\max}\leq2\norm{\left[H,V_n\right]}.
\end{equation}
The above discretization only achieves first-order accuracy, but one can improve this by switching to a higher-order scheme, which can reduce the cost of fault-tolerant implementation; see \cite[Appendix H]{Su2021} for details.

In the following, we evaluate this bound for a model of clustered Hamiltonians acting on $N$ sites:
\begin{equation}
    H:=\sum_{\mathcal{K}}H_{\mathcal{K}}+\sum_{\mathcal{K}\neq\mathcal{L}}H_{\mathcal{K}:\mathcal{L}}
    =\sum_{\mathcal{K}}\sum_{k,k'\in\mathcal{K}}H_{k,k'}
    +\sum_{\mathcal{K}\neq\mathcal{L}}\sum_{k\in\mathcal{K},l\in\mathcal{L}}H_{k,l}.
\end{equation}
Here, each term from the Hamiltonian acts on at most two sites and the sites are further grouped into clusters. We use calligraphic capital letters such as $\mathcal{K}$ and $\mathcal{L}$ to denote the clusters, and use $k$ to denote an arbitrary single site within $\mathcal{K}$. Assuming that Hamiltonian terms are normalized $\norm{H_{k,l}}\leq 1$, 
we have
\begin{equation}
    \norm{f'}_{\max}\leq 4\Lambda_{\text{ind}},\qquad
    \Lambda_{\text{ind}}:=\max_{\mathcal{L}}\max_{l\in\mathcal{L}}\sum_{\mathcal{K}}\sum_{k\in\mathcal{K}}\norm{H_{k,l}}.
\end{equation}

Altogether, we have discretized \eq{backward} with error at most
\begin{equation}
    \sum_{x,j}\frac{1}{\sigma_{x,t_j}^2}
    \abs{\mathrm{Trace}\left(O_x \widetilde{\rho}_x(t_j)\right)
    -\sum_{m}a_{x,m}\cos(t_j\omega_{x,m}+\phi_x)}\frac{2t_j^2\Lambda_{\text{ind}}}{L}
    =\mathcal{O}\left(\frac{\lambda\Lambda_{\text{ind}}T}{L}\right),
\end{equation}
where
\begin{equation}
\label{eq:def_lambda}
    \lambda:=\lambda_0+\lambda_1,\qquad
    \lambda_0:=\sum_{x,j}\frac{t_j}{\sigma_{x,t_j}^2},\qquad
    \lambda_1:=\sum_{x,j,m}\frac{a_{x,m}t_j}{\sigma_{x,t_j}^2},\qquad
    T:=\max_{j}t_j.
\end{equation}
To achieve an accuracy of $\epsilon$, it suffices to choose
\begin{equation}
\label{eq:def_l}
    L=\mathcal{O}\left(\frac{\lambda\Lambda_{\text{ind}}T}{\epsilon}\right).
\end{equation}
We take $T$ and $L$ to be powers of two to simplify our circuit implementation.

We now consider the error due to the finite-digit truncation of the real parameters $\sigma_{x,t_j}$, $t_j$, $a_{x,m}$, and $\omega_{x,m}$. In general, the error in $\sigma_{x,t_j}$ can be bounded under certain continuity assumptions with respect to the argument $t_j$. Here, we take $\sigma_{x,t_j}\equiv\sigma$ to be constant to simplify the analysis. In our circuit implementation, the evolution time will be loaded onto a quantum register using the QROM approach of Ref.~\cite{Babbush18Encoding} as 
\begin{equation}
    \ket{t_{\log T-1}\ \cdots\ t_{1}\ t_0}
    \ket{t_{-1}\ t_{-2}\ \cdots\ t_{-\log K}},
\end{equation}
where we have again taken $K$ to be a power of two to simplify the implementation. Here, $T$ is the maximum possible time so $\log T$ bits suffice to represent the integer part of $t$. The length of the decimal part should be chosen large enough to represent the time sufficiently accurate. Specifically, for $\abs{t'-t}\leq 1/K$, we have
\begin{equation}
    \norm{e^{-it'H}-e^{-itH}}\leq\abs{t'-t}\norm{H}=\mathcal{O}\left(\frac{N\Lambda_{\text{ind}}}{K}\right),
\end{equation}
which implies
\begin{align}
    \abs{\int_0^{t'}ds\ f(t',s)-\int_0^{t}ds\ f(t,s)}
    &\leq\abs{\int_0^{t'}ds\ f(t',s)-\int_0^{t}ds\ f(t',s)}+\abs{\int_0^{t}ds\ f(t',s)-\int_0^{t}ds\ f(t,s)}\nonumber\\
    &=\mathcal{O}\left(\frac{N\Lambda_{\text{ind}}t}{K}\right),\nonumber\\
    \abs{g(t')-g(t)}&\leq\abs{t'-t}\max_{\tau}\abs{g'(\tau)}=\mathcal{O}\left(\frac{N\Lambda_{\text{ind}}}{K}\right)
\end{align}
for
\begin{align}
    f(t,s)&:=\mathrm{Trace}\left(O_x\left[e^{-i(t-s)H}V_ne^{i(t-s)H},e^{-itH}\ketbra{\psi_x}{\psi_x}e^{itH}\right]\right),\nonumber\\
    g(t)&:=\mathrm{Trace}\left(O_x e^{-itH}\ketbra{\psi_x}{\psi_x}e^{itH}\right).
\end{align}
Similarly, if $t_j'$ and $\omega_{x,m}'$ satisfy $\abs{t_j'-t_j}\leq 1/K$ and $\abs{\omega_{x,m}'-\omega_{x,m}}\leq1/K$, then
\begin{equation}
    \abs{\cos(t_j'\omega_{x,m}'+\phi_x)-\cos(t_j\omega_{x,m}+\phi_j)}\leq\abs{t_j'\omega_{x,m}'-t_j\omega_{x,m}}=\mathcal{O}\left(\frac{T+W}{K}\right),
\end{equation}
where $T:=\max_j t_j$ and $W:=\max_{x,m}\abs{\omega_{x,m}}$. 
The coefficients in the Hamiltonian can be approximately prepared using the coherent alias sampling approach also described in Ref.~\cite{Babbush18Encoding}. Using that approach with $\log K$ qubits for the inequality test, we have
\begin{equation}
    \abs{\frac{t_j'/\sigma'^2}{\lambda_0'}-\frac{t_j/\sigma^2}{\lambda_0}}\leq\frac{1}{KN_d},\qquad
    \abs{\frac{a_{x,m}'t_j'/\sigma_j'^2}{\lambda_1'}-\frac{a_{x,m}t_j/\sigma_j^2}{\lambda_1}}\leq\frac{1}{KN_xN_\omega N_d}.
\end{equation}

The total truncation error can now be bounded by
\begin{equation}
    \mathcal{O}\left(\frac{\lambda N\Lambda_{\text{ind}}}{K}
    +\frac{\lambda_0 N\Lambda_{\text{ind}}}{K}
    +\frac{\lambda_1(T+W)}{K}
    +\frac{N_d\lambda_0}{KN_d}
    +\frac{N_xN_\omega N_d\lambda_1}{KN_xN_\omega N_d}\right)
    =\mathcal{O}\left(\frac{\lambda (N\Lambda_{\text{ind}}+T+W)}{K}\right).
\end{equation}
To ensure that this error is at most $\epsilon$, it suffices to choose
\begin{equation}
    K=\mathcal{O}\left(\mathrm{poly}(\lambda,N,\Lambda_{\text{ind}},T,W,1/\epsilon)\right).
\end{equation}

\section{Alternative derivation of Hamiltonian derivatives via optimal control theory}\label{app:optimal_control}
To derive Eq.~\ref{eq:cost_function_derivative} through optimal control theory, we enforce the evolution of $\rho_x(t)$ by $H+H_x(t)$ variationally.
We introduce an auxiliary field $\kappa_x(t)$ as a Lagrange variable to enforce this condition, which transforms our cost function to
\begin{align}
    C[\bar{H},\bar{\rho},\bar{\kappa}]=&\sum_n\frac{(\bar{h}_n-h_n^{(0)})^2}{2\omega_n^2}+\sum_{x,t}\frac{1}{2\sigma^2_{x,t}}\Big(\mathrm{Trace}\big[\bar{\rho}_x(t)O_x\big]-S_x(t)\Big)^2\nonumber\\
    &+i\sum_x\int_0^{\infty}dt\,\mathrm{Trace}\Bigg[\bar{\kappa}_x(t)\bigg(\frac{\partial\bar{\rho}_x(t)}{\partial t}-i\big[\bar{H}+H_x(t),\bar{\rho}_x(t)\big]\bigg)\Bigg].
\end{align}
This is now a functional; in addition to the finite real values $\bar{h}_n$, it also takes as input any smooth matrix-valued functions $\bar{\rho}_x(t)$ and $\bar{\kappa}_x(t)$.
This implies that all dependence of $C$ on $H$ is explicit; the dependence of $\bar{\rho}_x(t)$ (and $\bar{\kappa}_x(t)$ will emerge by the principle of least action.
The solution to our problem is given again by the minimum of the functional $C$.
Taking a functional derivative $\frac{\delta C}{\delta\bar{\kappa}_x(t)}=0$ yields the Schr\"{o}dinger equation in its standard form
\begin{equation}
    \frac{\delta C}{\delta\bar{\kappa}_x(t)}=0\rightarrow \frac{\partial\bar{\rho}_x(t)}{\partial t}=i\big[\bar{H}+H_x(t),\bar{\rho}_x(t)\big],
\end{equation}
which yields the solution in the main text: $\bar{\rho}_x(t)=\bar{U}_x(t,0)\rho_x\bar{U}_x^{\dag}(t,0)$,
Taking the functional derivative with respect to $\bar{\rho}_x(t)$ and setting this equal to $0$ yields an update rule for $\bar{\kappa}_x(t)$
\begin{equation}
    \frac{\partial\bar{\kappa}_x(t)}{\partial t}=-i\big[\bar{\kappa}_x(t),H]-i\sum_{t'}\frac{1}{\sigma_{x,t'}^2}\delta(t-t')\Big(\mathrm{Trace}\big[\bar{\rho}_x(t')O_x\big]-S_x(t')\Big)O_x,\;\;\;\kappa_x(+\infty)=0.
\end{equation}
This takes the form of an external field $\bar{\kappa}_x(t)$ that propagates back in time and is perturbed in a non-unitary way by each measurement $S_x(t')$ that does not completely match the predicted $\bar{S}_x(t')$.
Substituting in the solution for $\bar{\rho}_x(t)$ yields a solution for $\bar{\kappa}_x(t)$
\begin{equation}
    \kappa_x(t)=i\sum_{t'>t}\frac{1}{\sigma_{x,t'}^2}\bar{U}^{\dag}_x(t,t')O_x\bar{U}_x(t,t')\Big(\mathrm{Trace}\big[\bar{\rho}_x(t)O_x\big]-S_x(t)\Big).\label{eq:kappa_solution}
\end{equation}
To recover the update rule, we then take the partial derivative of $C$ with respect to the parameters $h_n$ and set this to zero
\begin{equation}
    \frac{\partial C}{\partial \bar{h}_n}=\frac{(\bar{h}_n-h_n^{(0)})}{\omega_n^2}+\sum_x\int_0^{\infty}dt\,\mathrm{Trace}\Big[\bar{\kappa}_x(t)\big[V_n,\bar{\rho}_x(t)\big]\Big].
\end{equation}
Substituting in Eq.~\ref{eq:kappa_solution} yields Eq.~\ref{eq:cost_function_derivative} as required.
Following a similar procedure to take second-order derivatives of $C$ with respect to $\bar{\rho}_x(t)$ and $\bar{\kappa}_x(t)$ yields Eq.~\ref{eq:cost_function_second_derivative} after some rearrangement.

\end{document}